\documentclass[a4paper,11pt]{article}
\usepackage{jheppub}
\usepackage{graphicx}
\usepackage{amssymb}
\usepackage{slashed}
\usepackage{feynmf}
\usepackage{braket}
\usepackage{empheq}
\usepackage{dcolumn}
\usepackage{color}
\usepackage{comment}
\usepackage[font=small]{caption}
\usepackage[font=small]{subcaption}
\newcommand{\be}{\begin{eqnarray}}
\newcommand{\ee}{\end{eqnarray}}
\newcommand{\nn}{\nonumber}
\newcommand{\f}{\frac}
\newcommand{\p}{\partial}

\usepackage{tikz}
\usetikzlibrary{arrows.meta,arrows}

\title{\boldmath Perturbative soft graviton theorems in de Sitter spacetime}
\author[a]{Divyesh N. Solanki,}
\author[b,c]{Pratik Chattopadhyay}
\author[a]{and Srijit Bhattacharjee}
\affiliation[a]{Indian Institute of Information Technology Allahabad (IIITA), Devghat, Jhalwa, Prayagraj-211015, Uttar Pradesh, India.}
\affiliation[b]{School of Physics, The University of Electronic Science and Technology of China, No.2006, Xiyuan Avenue, West Hi-Tech Zone, Chengdu, Sichuan, P.R.China, Post Code: 611731}
\affiliation[c]{Tsung-Dao Lee Institute, Shanghai Jiao Tong University, Shanghai, China}
\emailAdd{divyeshsolanki98@gmail.com}
\emailAdd{pratikpc@gmail.com}
\emailAdd{srijuster@gmail.com}

\abstract{We consider soft graviton scattering for a theory where Einstein's gravity is minimally coupled to a scalar field in the presence of a cosmological constant, i.e. in a background de Sitter space. Employing a perturbative expansion in a small cosmological constant, we compute leading, subleading and sub-subleading corrections to Weinberg's soft graviton amplitude for the tree-level scatterings in the static patch of de Sitter space. We observe similar universal features of the soft graviton amplitude as found in \cite{Sayali_Diksha} for the soft photons.}

\begin{document}
\maketitle
\flushbottom

\section{Introduction}

The infrared properties of gauge theories have been of great interest for a long time, particularly, the soft factorization where the scattering amplitude for processes involving the emission of soft particles is related to the amplitude without soft particles by a universal factor, known as the soft factor, which depends solely on the momentum of the scatterer and the energy of the soft particle \cite{Bloch_1937, Gell-Mann_Goldberger, Low_1954, Low_1958, Weinberg_1965, Gross_Jackiw, Jackiw_1968, White_2011}. This soft factorization property of gauge theories, known as soft theorems, later realized as Ward identities of asymptotic symmetries \cite{Temple_He, Laddha_2015, Daniel_Kapec, He_Lysov_Mitra_Strominger, Lysov_Pasterski_Strominger, Campiglia_Laddha_2016, Kapec_Lysov_2014, Campiglia_Laddha_2014}. Another key aspect of infrared physics of gauge theories is the memory effect \cite{Zeldovich-Polnarev, BT,Strominger_Zhiboedov, Bieri}, which has a trivial mathematical relationship with the soft theorems given by the Fourier transform \cite{Strominger_Zhiboedov, Pasterski_Strominger_Zhiboedov}. The memory effect can also be understood as the asymptotic symmetry transformation. The passage of radiation through the future null infinity ($\mathcal{I}^+$) produces a transition in the vacuum that is equivalent to the asymptotic symmetry transformation \cite{Strominger_Zhiboedov}. All such interconnections among three different theories are summarized in the infrared triangle \cite{Strominger}. There exists many such infrared triangles for the tree-level scattering processes. The loop effects introduce additional terms to the soft factor that depend on the logarithm of the soft energy, resulting in the logarithmic soft theorems \cite{Laddha_Sen_2018, SS_2019, SSS_2020}, which are related to the tail memory effects \cite{Laddha_Sen_2019, SS_2022}. The asymptotic symmetries corresponding to the logarithmic soft theorems have been discovered recently in \cite{Campiglia_Laddha_2019, Agrawal_Donnay_2024, Choi_Laddha_Puhm_1}, and the complete description of the infrared triangle is given in \cite{Choi_Laddha_Puhm_2, Choi_Laddha_Puhm_3}.

The early and late time limits of our universe are well-approximated by the de Sitter space, where in the context of early universe, examining the soft behavior of in-in correlators and wavefunction coefficients has drawn considerable interest \cite{Armstrong_2022, Maldacena_2003, Paolo_2012, Kurt_2014, Kundu_2015, Paolo_2003, Valentin_2012, Kundu_2016, Shukla_2016, Chowdhury_2024}. Higher-point correlators can be related to conformal transformations of lower-point correlators, and certain three-point correlators can be deduced by taking the soft limit of four-point correlators. In the context of late-time universe, where the cosmological constant is extremely small, the study of cosmological effects on soft amplitudes and memory observables have been explored in \cite{Sayali, Sayali_Diksha} using a perturbative approach. The soft theorems in the de Sitter space have been derived only recently from the Ward identities of near cosmological horizon symmetries \cite{Mao_Zhou, Mao_Zhang}, which is consistent with the flat space limit. The memory effects have also been studied extensively in the de Sitter space to understand cosmological implications on these observables \cite{Bieri_2016, Chu_2017, Chu_2021, Tolish_Wald_2016}. Further, the relationship between gravitational wave memory and BMS-like supertranslations has been investigated in the static patch of the de Sitter space in \cite{Hamada_2017}. In the context of present universe where the thermal effects prevail, the finite temperature effects on soft theorems and memory effects have been explored in \cite{SB_2023}.

Recently, the soft photon theorems in de Sitter space was studied using the perturbative approach \cite{Sayali_Diksha}. The perturbative corrections to the flat space soft photon theorem in $d$ dimensions are obtained by analyzing the soft behavior of the perturbative $\mathcal{S}$-matrix defined in the static patch \cite{Emil_2021}, and the universality of these corrections was analyzed. This analysis has not yet been done for the soft graviton theorems in de Sitter space. It is interesting to see if the universality of the soft graviton theorems in de Sitter space is consistent with that of the soft photon theorems. In this paper, we use the same perturbative approach to derive soft graviton theorems in four dimensions in the de Sitter space. Our setup is as follows: We confine scattering processes to a compact region of size $R$ inside the static patch, which is much smaller than the de Sitter curvature length $l$ \cite{Sayali_Diksha, Mandal_Banerjee} so that gravitational effects can be treated perturbatively, and much larger than the Planck length so that the gravitational interactions become weak as particles approach the boundaries. The graviton detector is also placed in the same region. The boundaries of this region can be reached in a finite time which is assumed to be much larger than the interaction time scale. Thus, the particles are asymptotically decoupled \cite{Hijano_Neuenfeld}, which means they are far away from each other and non-interacting at early and late times. Thus, the early and late time Cauchy surfaces are the Hilbert spaces of the incoming and outgoing particles, respectively. Our results are valid in the regime $\sqrt{G}<<R<<l$.
Our approach is different than that of \cite{Mao_Zhou}, where the asymptotic in- and out-states are defined on the cosmological horizons and soft graviton theorems are obtained via Ward identities.

This paper is organized as follows: In section (\ref{dS_spacetime_sec}), we briefly describe the geometry of the de Sitter space and define appropriate coordinates for our analysis. We review the scalar fields in the de Sitter background in section (\ref{dS_review_sec}), which have been studied in details in \cite{Sayali_Diksha} for generic dimensions $d$. Next, we provide linearized Einstein field equation in the de Sitter background in section (\ref{Linearized_sec}) and obtain mode solutions in section (\ref{mode_expansion_sec}). By defining the inner product of the modes, we check their orthogonality and consider only the orthogonal set of modes in our analysis. Then, we derive the LSZ formula for gravitons by utilizing the inner product of modes in section (\ref{LSZ_graviton_sec}). In section (\ref{S-matrix_sec}), we define the perturbative $\mathcal{S}$-matrix in the static patch of the de Sitter space as in \cite{Sayali_Diksha}, which differs from the perturbative $\mathcal{S}$-matrix defined in global de Sitter space in \cite{Marolf_2013}. Utilizing the LSZ formula, we compute the $\mathcal{S}$-matrix for the scattering process depicted in Fig. (\ref{soft_graviton_fig}) in section (\ref{soft_graviton_theorem_sec}). We compute perturbative corrections\footnote{By perturbative corrections, we refer to the $\mathcal{O}(l^{-1})$ and $\mathcal{O}(l^{-2})$ corrections to the flat space soft factor, which should not be confused with leading, subleading, and sub-subleading corrections. As analyzed in section (\ref{subleading_sec}), the $\mathcal{O}(l^{-2})$ corrections include terms of order $1/k^3$ and $1/k^2$, which do not contribute at the same order. The former is more leading than the latter. Thus, $\mathcal{O}(l^{-2})$ corrections split into subleading and sub-subleading corrections.} to the flat space soft factor and deduce leading, subleading, and sub-subleading corrections in the subsequent subsections (\ref{leading_sec}) and (\ref{subleading_sec}). In (\ref{Internal_line_sec}), we compute contributions from the second diagram (Fig. (\ref{soft_internal_fig})) where a soft graviton is attached to an internal line. We do consistency checks of our results in (\ref{consistency_checks}). Finally, we conclude by summarizing our results and providing some remarks in section (\ref{discussion_sec}).

\section{De Sitter spacetime}
\label{dS_spacetime_sec}
De Sitter (dS) spacetime is the simplest non-trivial spacetime with constant positive scalar curvature. The Penrose diagram of such a spacetime is a square, depicted in Fig. (\ref{dS_Penrose}).  One cannot chart the full spacetime with a single coordinate system, but instead can specify some local coordinates in the static patch of dS. Let us consider a static coordinate system $(t,r,\theta^A)$ in which the dS metric takes the following form:
\begin{equation}
    ds^2 = -\left(1-\dfrac{r^2}{l^2} \right) dt^2 + \left(1-\dfrac{r^2}{l^2} \right)^{-1} dr^2 + r^2 d\Omega_2^2,
\end{equation}
where $0\leq r \leq l$, and $l$ is the radius of curvature of the de-Sitter space. The point $r=l$ is not a real singularity but a coordinate singularity beyond which a static coordinate system cannot be extended.
\\~\\
We can also view the de Sitter space as an embedding of the hyperboloid,
\begin{equation}
\label{hyperboloid}
    \eta_{AB}X^AX^B=l^2,
\end{equation}
in the ambient Minkowski space $\eta_{AB}$, where $X^A$ are the Cartesian coordinates and $A$ runs from $0$ to $4$. As we shall see, the metric of such a spacetime is conformally equivalent to the flat space. This is best seen as a stereographic projection of the hyperboloid to the flat Minkowski space. To do so, we use the stereographic mapping as given in \cite{Aldrovandi_1995, Sayali_Diksha}. 
\begin{equation}
X^{\mu}=\dfrac{x^{\mu}}{1+x^2/4l^2}, \ \ \mu=0,1,2,3.
\end{equation}
Where, $x^2=\eta_{\mu\nu}x^{\mu}x^{\nu}$, and the last coordinate $X^4$ is fixed by the relation (\ref{hyperboloid}). In the stereographic coordinates $x^\mu$, the line element $ds^2=\eta_{AB}dX^AdX^B$ takes the form $ds^2=g_{\mu\nu}dx^{\mu}dx^{\nu}$ with 
\be 
g_{\mu\nu}=\Omega^2\eta_{\mu\nu}, ~~\Omega=\frac{1}{1+x^2/4l^2}.
\ee 

\begin{figure}
\centering
\begin{tikzpicture}
\draw[black, very thick] (-2.5,2.5) to node[shift={(0,0.3)}, sloped] {$\mathcal{I}^+$} (2.5,2.5);
\draw[black, very thick] (-2.5,-2.5) to node[shift={(0,-0.3)}, sloped] {$\mathcal{I}^-$} (2.5,-2.5);
\draw[black, very thick] (-2.5,2.5) to node[shift={(-1.1,0)}] {North Pole} (-2.5,-2.5);
\draw[black, very thick] (2.5,2.5) to node[shift={(1.1,0)}] {South Pole} (2.5,-2.5);
\draw[black, very thick] (-2.5,2.5) to node[shift={(-1,1.6)}] {$H^+$} (2.5,-2.5);
\draw[black, very thick] (2.5,2.5) to node[shift={(-1,-1.5)}] {$H^-$} (-2.5,-2.5);
\draw[fill=orange] (-2.5,2.5) -- (0,0) -- (-2.5,-2.5) -- cycle;
\draw[black, very thick,fill=lightgray] (-2.5,1.5) to [out=-45, in=45,looseness=2] node[shift={(-0.8,0)}] {$R$} (-2.5,-1.5);
\end{tikzpicture}
\caption{Penrose diagram of the de Sitter space. The gray region in the static patch is a compact region $R$ where scattering processes are confined.}
\label{dS_Penrose}
\end{figure}

\noindent Where, $\eta_{\mu\nu}$ is the Minkowski metric. The advantage of using the above coordinate system is that the metric is conformally flat. We must note that there is a coordinate singularity at $x^2 = -4l^2$, which however does not affect our analysis. Because, we confine a scattering process to a compact region of size $R$ in the static patch (Fig. (\ref{dS_Penrose})), where $R<<l$. Therefore, all the points in region $R$ have $x^\mu<<l$ for every component ``$\mu$", and hence the above metric takes the following form up to $\mathcal{O}(l^{-2})$:
\be 
\label{metric}
g_{\mu\nu}\approx\Big(1-\frac{x^2}{2l^2}\Big)\eta_{\mu\nu}.
\ee

\section{Scalar fields in de Sitter background: Review}
\label{dS_review_sec}
\subsection{Scalar modes}
In this section, we briefly review the derivation of the scalar modes in $d=4$, which is explained in \cite{Sayali_Diksha} for arbitrary $d$ dimensions in details. The free scalar field equation is given as
\be
\label{scalareom}
[\nabla^2-m^2]\phi =0,\ee
where $\nabla^2$ is the d'Alembertian operator in the de Sitter spacetime and $m$ is the mass of the scalar field $\phi$. Considering corrections up to $\mathcal{O}(l^{-2})$, this equation has the form 
\be \label{sm}
[\nabla^2-m^2]\phi = \left(1+\frac{x^2}{2l^2}\right)\Box\phi -\frac{1}{l^2}x\cdot\partial\phi-m^2\phi=0,
\label{scalareom1}
\ee
where $\Box$ is the flat space d'Alembertian operator. We follow the deduction given in \cite{Sayali_Diksha}, where the solution for the scalar field equation in embedding space coordinates $X^A$ is given by  
\be 
g_{p}= {\mathcal{C}_l}\left(\f{X^A\xi^B\eta_{AB}}{l}+i\epsilon\right)^{-\Delta},
\label{gp_general}
\ee
where $\mathcal{C}_l$ is a normalization constant. The vector $\xi^A$ and $\Delta$ satisfy $\eta_{AB}\xi^A \xi^B = 0$ and 
$$\Delta^2-3\Delta+m^2 l^2 = 0,$$
respectively. The $\Delta$ has two solutions
$$\Delta_{\pm}=\dfrac{1}{2}\big(3\pm \sqrt{9-4m^2l^2} \big),$$
where $\Delta_+$ corresponds to outgoing modes $(e^{ip\cdot x})$ and $\Delta_-$ corresponds to incoming modes $(e^{-ip\cdot x})$ in the flat space. 
It is also customary to mention that we can parametrize $\xi^A=(\frac{p^{\mu}}{m},1)$ such that $\eta_{\mu\nu}p^\mu p^\nu = -m^2$.

Using the correct limiting behaviour for the scalar function $g_p$ in the limit $l\rightarrow \infty$, the function $\mathcal{C}_l$ is obtained as
\be \label{C}\mathcal{C}_l =e^{-m\pi l + \frac{3i\pi}{2}+\f{9\pi}{8ml}}\Big[1 - \f{1}{m^2 l^2}\ \Big].
\ee

Substituting $\mathcal{C}_l$, $\Delta_+$, $\xi^A$, and $X^A$ in Eq. (\ref{gp_general}) and keeping terms up to $\mathcal{O}(l^{-2})$, we obtain \cite{Sayali_Diksha} 
\begin{multline}
\label{gp_modes}
g_p(x)=\f{e^{ip\cdot x}}{\sqrt{2E_p}}\Big[1+\f{3(p\cdot x)}{2ml}+\f{imx^2}{2l}+\f{i(p\cdot x)^2}{2ml}- \f{1}{m^2 l^2}-\f{9i(p\cdot x)}{8m^2l^2} + \f{3x^2}{4l^2} +\f{15(p\cdot x)^2}{8m^2l^2}\\
+\f{\ i(p\cdot x)\ x^2}{l^2}+\f{13i(x\cdot p)^3}{12m^2l^2}-\f{(p\cdot x)^4}{8m^2l^2}-\f{(p\cdot x)^2 x^2}{4l^2}-\f{m^2 x^4}{8l^2}\Big].
\end{multline}
It can be checked that $g_p$ solves \eqref{scalareom1} with the on-shell condition $$p^2=-E_p^2+{|\vec{p}|}^2=-m^2,$$
where $E_p$ is the energy of the scalar field modes. The modes are orthogonal which can be seen from the inner product defined as \cite{Sayali_Diksha}
\begin{equation}
\label{scalar_inner_product}
    \left( g_p, g_q \right) = -i \int d^3x \sqrt{-g} \ e^{-\epsilon\frac{|\vec{x}|}{R}} \big(g^*_p(x) \nabla^t g_q(x) - g_q(x) \nabla^t g^*_p(x) \big),
\end{equation}
where the exponential damping factor is introduced to remove the boundary terms. Substituting the modes (\ref{gp_modes}) in the above expression, one can verify that
\begin{equation}
    \left( g_p, g_q \right) = (2\pi)^3 \delta^{(3)}(p-q).
\end{equation}
Thus, we can now mode expand the scalar field as 
\be
\phi(x)=\int \f{d^{3}p}{(2\pi)^{3}} \ [\ a_p\ g_p(x)\ +\ a^\dagger_p\ g^*_p (x)\ ],\nn
\ee
where $a_p$ and $a_p^{\dagger}$ are the annihilation and creation operators, respectively.

\subsection{Alternate class of scalar modes}
In the above section, we deduced the scalar field modes by demanding that they obey the correct flat space limit. However, as shown in \cite{Sayali_Diksha}, these are not the only modes to satisfy this limit. The following modes, namely 
\be 
\label{hp_modes}
h_p=\frac{e^{ip\cdot x}}{\sqrt{2E_p}}\bigg[1+\frac{3x^2}{4l^2}-\frac{i(p\cdot x) x^2}{4l^2}-\frac{i(p\cdot x)^3}{6m^2l^2}+\frac{c}{l^2}\bigg(\frac{3p\cdot x}{2m}+\frac{imx^2}{2}+\frac{i(p\cdot x)^2}{2m}\bigg)\bigg]
\ee 
with 
\be 
p^2=-m^2+\frac{6}{l^2}
\ee 
also satisfy the scalar field equation of motion, with an arbitrary constant $c$. In the $l\to \infty$ limit, the solution behaves correctly as plane waves. In this paper, we use the modes $g_p$ to compute the $\mathcal{S}$-matrix because the modes $h_p$ are not orthogonal.
 
\subsection{Propagator for the scalar field}
The scalar field propagator up to $\mathcal{O}(l^{-2})$ in the de Sitter space has been derived in \cite{Sayali_Diksha} for arbitrary dimensions $d$. We thus do not run over the full computation, but instead briefly mention the steps required to derive it and then present the final result. The differential equation for the propagator can be stated as 
\be\label{xeq}
[\nabla_x^2- m^2]\ D(x,y)=i \frac{\delta^{(4)}(x-y)}{\sqrt{-g}}.
\ee
Expanding the covariant derivative, we get 
\be 
\label{propeom}
\left[\left(1+\frac{x^2}{2l^2}\right)\Box_x-m^2-\frac{1}{l^2}x\cdot\p_x\right]D(x,y) = i \frac{\delta^{(4)}(x-y)}{\sqrt{-g}}.
\ee
The solution $D$ can be expanded as follows \cite{Sayali_Diksha}
\be
{D}(x,y) = {D}_0(x,y)+ \delta D(x,y),
\ee
where
\be 
{D}_0(x,y)=-i\int \frac{d^4p}{(2\pi)^4}\frac{e^{ip\cdot(x-y)}}{p^2+m^2-i\epsilon}
\ee 
is the usual flat space propagator and $\delta {D}(x,y)$ denotes perturbative corrections. The Fourier expansion of $\delta D$ is
\be
\delta D(x,y) = -i\int \frac{d^4p}{(2\pi)^4} e^{ip\cdot(x-y)} \delta\Tilde{D}(p,y).
\ee
Substituting the decomposition of the propagator in terms of $D_0$ and $\delta D$ in Eq. (\ref{propeom}), and performing integration by parts in $p$, the full propagator up to $\mathcal{O}(l^{-2})$ takes the following form
\begin{align}
\label{asymmetric_prop}
D(x,y)=-i\int\frac{d^4p}{(2\pi)^4}\frac{e^{ip\cdot(x-y)}}{p^2+m^2}\Big[ & 1+\frac{4m^4}{l^2(p^2+m^2)^3}+\frac{2m^2(1-ip\cdot y)}{l^2(p^2+m^2)^2}\nn\\
&+\frac{m^2y^2-2ip\cdot y+4}{2l^2(p^2+m^2)}+\frac{y^2}{2l^2}\Big].
\end{align}
It is not hard to check that $D(x,y)$ also satisfies Eq. (\ref{propeom}) when $x$ and $y$ are interchanged. Though the above form of the propagator is not explicitly symmetric in $x$ and $y$, one can bring it to an explicitly symmetric form using integration by parts \cite{Sayali_Diksha}. In the following section, we discuss the symmetric form of the propagator.

\subsubsection{Symmetric form of Feynman propagator}\label{fnp}
We can write the symmetric form of the Feynman propagator using the orthogonal set of modes $g_p$ as follows: \cite{Sayali_Diksha} 
\begin{equation}
    D(x,y)=-i\int\dfrac{d^4p}{(2\pi)^4} \dfrac{g_p(x)g^*_p(y)}{p^2+m^2-i\epsilon}.
\end{equation}
Substituting Eq. (\ref{gp_modes}) into the above expression, and keeping terms up to $\mathcal{O}(l^{-2})$, the Feynman propagator can be decomposed into the flat space propagator and corrections of the order  $\mathcal{O}(l^{-1})$ and $\mathcal{O}(l^{-2})$, respectively. Thus, we have
\be
\label{symmetric_prop}
 D(x,y) =\  D^{(0)}(x,y)+ D^{(1)}(x,y) + D^{(2)}(x,y).
\ee
Now, we state the expressions of the flat as well as correction terms in $d=4$. This has already been explicitly shown in \cite{Sayali_Diksha} and we just restate it here so that we can use it in our calculations. 
\begin{align}
{D}^{(0)}&= -i \int \f{d^4 p}{(2\pi)^4}\f{e^{ip\cdot(x-y)}}{p^2+m^2},\nn\\
 D^{(1)}&=  -i  \int \f{d^4 p}{(2\pi)^4}\f{e^{ip\cdot(x-y)}}{p^2+m^2}\ \Big[ \frac{3 p\cdot x}{2 l m}+\frac{3 p\cdot y}{2 l m}+\frac{i (p\cdot x)^2}{2 l m}-\frac{i (p\cdot y)^2}{2 l m}+\frac{i m x^2}{2 l}-\frac{i m y^2}{2 l}\Big]=0,\nn\\
 D^{(2)}&=  -i  \int \f{d^4 p}{(2\pi)^4}\f{e^{ip\cdot(x-y)}}{p^2+m^2}\ \Big[ \frac{15 (p\cdot x)^2}{8 l^2 m^2}+\frac{15(p\cdot y)^2}{8 l^2 m^2}+\frac{ 3i (p\cdot x)^2 (p\cdot y)}{4 l^2 m^2}+\frac{9 (p\cdot x) (p\cdot y)}{4 l^2 m^2}\nn\\
 &-\frac{3i (p\cdot x) (p\cdot y)^2}{4 l^2 m^2}+\frac{ 13i (p\cdot x)^3}{12 l^2 m^2}-\frac{ 9i (p\cdot x)}{8 l^2 m^2}+\frac{ 9i (p\cdot y)}{8 l^2 m^2}-\frac{13i (p\cdot y)^3}{12 l^2 m^2} + \frac{3i x^2 (p\cdot y)}{4 l^2}\nn\\
 &+\frac{i x^2 (p\cdot x)}{ l^2}-\frac{ 3iy^2 (p\cdot x)}{4 l^2}-\frac{i y^2 (p\cdot y)}{ l^2}+\frac{3x^2}{4 l^2}+\frac{3 y^2}{4 l^2}+\frac{(p\cdot x)^2 (p\cdot y)^2}{4 l^2 m^2}-\frac{(p\cdot x)^4}{8 l^2 m^2}\nn\\
 &-\frac{(p\cdot y)^4}{8 l^2 m^2}-\frac{m^2 x^4}{8 l^2}+\frac{m^2 x^2 y^2}{4 l^2}-\frac{m^2 y^4}{8 l^2}+\frac{x^2 (p\cdot y)^2}{4 l^2}-\frac{x^2 (p\cdot x)^2}{4 l^2}+\frac{y^2 (p\cdot x)^2}{4 l^2}-\frac{y^2 (p\cdot y)^2}{4 l^2}\Big].
\end{align}
As is obvious from the expressions, the propagator stated above is manifestly symmetric in $x$ and $y$. A comparison of this propagator with the asymmetric one is already done in \cite{Sayali_Diksha} and so we do not re-do it here. We would mention however that $\mathcal{O}(l^{-1})$ correction $D^{(1)}$ vanishes. This is because we can express the variable $x$ as $-i\partial_p$ and convert it into a $y$ dependent piece by doing an integration by parts in $p$. Then, it is not hard to show that these manipulations render $D^{(1)}$ to be zero. We note this is consistent with the absence of a $1/l$ piece in the asymmetric form of the propagator derived in the previous section. Another thing we want to point out is that although the $\mathcal{O}(l^{-2})$ correction to the flat space propagator does not seem to match with that in the previous section, we can easily do integration by parts in $p$ and match it thereafter. Thus, all in all, the symmetric form of the Feynman propagator is consistent with its asymmetric form.\footnote{It is worthwhile to mention that the consistency of the two forms of the propagator means one form can be transformed into another by doing integration by parts in $p$.}

\section{Gravitons in de Sitter background}
In this section, we briefly study gravitons in the de Sitter space. We derive linearized Einstein field equation and determine its mode solutions. We define an inner product of the modes and check their orthogonality. By inverting the inner product expression, we derive the LSZ formula for gravitons.

\subsection{Linearized field equation}
\label{Linearized_sec}
To deduce the linearized Einstein field equation in the de Sitter background, a general spacetime metric can be expanded as \cite{Date_Hoque_2016, Mao_Zhou}
\be
g_{\mu\nu} = \Bar{g}_{\mu\nu} + \kappa h_{\mu\nu},
\ee
and its inverse as
\be
g^{\mu\nu} = \Bar{g}^{\mu\nu} - \kappa h^{\mu\nu} + \mathcal{O}(\kappa^2),
\ee
where $\kappa=\sqrt{32\pi G}$, and $\Bar{g}_{\mu\nu}$ is the background de Sitter metric. The indices of $h_{\mu\nu}$ are raised and lowered by $\Bar{g}_{\mu\nu}$.\\
Let us now expand the Christoffel symbols up to the first order correction $\mathcal{O}(\kappa)$ as:
\begin{equation}
    \Gamma^{\alpha}_{\mu\nu} = \Bar{\Gamma}^{\alpha}_{\mu\nu} + \dfrac{\kappa}{2} \Bar{g}^{\mu\nu} \left( \Bar{\nabla}_\mu h_{\nu\beta} + \Bar{\nabla}_\nu h_{\mu\beta} - \Bar{\nabla}_\beta h_{\mu\nu} \right),
\end{equation}
where $\Bar{\Gamma}^\alpha_{\mu\nu}$ and $\Bar{\nabla}_\alpha$ are respectively the Christoffel symbols and covariant derivatives compatible with the de Sitter metric. Now, we determine the Riemann curvature tensor up to $\mathcal{O}(\kappa)$:
\begin{equation}
\label{Riemann_dS}
    \mathcal{R}_{\mu\nu\alpha}^{\ \ \ \ \beta} = \Bar{\mathcal{R}}_{\mu\nu\alpha}^{\ \ \ \ \beta} \ + \dfrac{\kappa}{2} \Bar{\nabla}_{\nu} \left(\Bar{\nabla}_\mu h^{\beta}_{\alpha} + \Bar{\nabla}_\alpha h^{\beta}_{\mu} - \Bar{\nabla}^\beta h_{\mu\alpha} \right)
    - \dfrac{\kappa}{2} \Bar{\nabla}_{\mu} \left(\Bar{\nabla}_\nu h^{\beta}_{\alpha} + \Bar{\nabla}_\alpha h^{\beta}_{\nu} - \Bar{\nabla}^\beta h_{\nu\alpha} \right),
\end{equation}
where $\Bar{\mathcal{R}}_{\mu\nu\alpha}^{\ \ \ \ \beta}$ is the Riemann curvature tensor of the de Sitter space.
To obtain the Einstein tensor, we require the Ricci tensor and the Ricci scalar. Contracting the indices $\nu$ and $\beta$ in (\ref{Riemann_dS}) yields the Ricci tensor
\begin{equation}
\label{Riccit_dS}
    \mathcal{R}_{\mu\alpha} = \Bar{\mathcal{R}}_{\mu\alpha}
    +\dfrac{\kappa}{2} \Bar{\nabla}_{\nu} \left(\Bar{\nabla}_\mu h^{\nu}_{\alpha} + \Bar{\nabla}_\alpha h^{\nu}_{\mu} - \Bar{\nabla}^\nu h_{\mu\alpha} \right)
    -\dfrac{\kappa}{2} \Bar{\nabla}_\mu \Bar{\nabla}_\alpha h,
\end{equation}
where $h=\Bar{g}^{\mu\nu}h_{\mu\nu}$ is the trace of $h_{\mu\nu}$. Similarly, contracting $\mu$ and $\alpha$ in the above expression yields the Ricci scalar
\begin{equation}
\label{Riccis_dS}
    \mathcal{R} = \Bar{\mathcal{R}} - \kappa h^{\mu\alpha} \Bar{\mathcal{R}}_{\mu\alpha}
    +\dfrac{\kappa}{2} \Bar{\nabla}_{\nu} \left(\Bar{\nabla}_\mu h^{\mu\nu} + \Bar{\nabla}^\mu h^{\nu}_{\mu} - \Bar{\nabla}^\nu h \right)
    -\dfrac{\kappa}{2} \Bar{\nabla}_\mu \Bar{\nabla}^\mu h.
\end{equation}
Let us now consider the vacuum Einstein field equation with the cosmological constant term.
\begin{equation}
    E_{\mu\nu}=\mathcal{R}_{\mu\nu} - \dfrac{1}{2} g_{\mu\nu} \mathcal{R} + \Lambda g_{\mu\nu} = 0,
\end{equation}
where $\Lambda=\dfrac{3}{l^2}$ for the de Sitter space. Substituting Eqs. (\ref{Riccit_dS}) and (\ref{Riccis_dS}) into the above expression, we obtain the linearized form of the Einstein field equation
\begin{multline}
    E_{\mu\nu} = \dfrac{1}{2} \left(\Bar{\nabla}_\tau \Bar{\nabla}_\mu h^\tau_\nu + \Bar{\nabla}_\tau \Bar{\nabla}_\nu h^\tau_\mu - \Bar{\nabla}_\mu \Bar{\nabla}_\nu h - \Bar{\nabla}^2 h_{\mu\nu} \right) \\- \dfrac{1}{2}\Bar{g}_{\mu\nu} \left(\Bar{\nabla}_\alpha \Bar{\nabla}_\beta h^{\alpha\beta} - \Bar{\nabla}^2 h \right) - \dfrac{3}{l^2}h_{\mu\nu} + \dfrac{3}{2l^2} \Bar{g}_{\mu\nu}h = 0,
\end{multline}
where the following relations for the de Sitter space are used: $\Bar{\mathcal{R}}_{\mu\nu}=\dfrac{3}{l^2} \Bar{g}_{\mu\nu}$, $\Bar{\mathcal{R}}=\dfrac{12}{l^2}$.\\
The linearized equation obtained above is invariant under the gauge transformations \cite{Mao_Zhou}
\be
h_{\mu\nu} \to h_{\mu\nu} + \Bar{\nabla}_{\mu}\xi_{\nu} + \Bar{\nabla}_{\nu}\xi_{\mu}.
\ee
Applying the de Donder gauge condition, $\Bar{\nabla}_\mu h^{\mu\nu} = \dfrac{1}{2}\Bar{\nabla}^\nu h$, the linearized field equation reduces to
\begin{equation}
\bar{\nabla}^2h_{\mu\nu} - \dfrac{1}{2}\bar{g}_{\mu\nu} \bar{\nabla}^2 h - \dfrac{2}{l^2}h_{\mu\nu} - \dfrac{1}{l^2}\bar{g}_{\mu\nu} h = 0.
\end{equation}
We can fix the gauge $h=0$ as in the flat space as follows \cite{Higuchi}. First, we take the trace of the above expression with $h_{\mu\nu}$ replaced by $h'_{\mu\nu}$ as
\begin{equation}
\label{h'_eqn}
\left(\bar{\nabla}^2 + \dfrac{6}{l^2} \right) h' = 0.
\end{equation}
Let us now define
\begin{equation}
h_{\mu\nu} = h'_{\mu\nu} + \dfrac{1}{6/l^2} \bar{\nabla}_\mu \bar{\nabla}_\nu h',
\end{equation}
which is traceless from Eq. (\ref{h'_eqn}), and satisfies the de Donder gauge condition. Thus, the linearized field equation reduces to
\begin{equation}
\label{linearized_Einstein_dS}
\left(\bar{\nabla}^2 - \dfrac{2}{l^2} \right) h_{\mu\nu} = 0,
\end{equation}
and the de Donder gauge reduces to the transverse-traceless gauge
\begin{equation}
\label{TT-gauge}
\bar{\nabla}_\mu h^{\mu\nu} = 0 = h.
\end{equation}
In the following sections, we will drop the bar notations for the de Sitter background.

\subsection{Mode solutions}
\label{mode_expansion_sec}
In this section, we find the mode expansion for gravitons in the de Sitter space. We first write the linearized field equation (\ref{linearized_Einstein_dS}) in the stereographic coordinates up to $\mathcal{O}(l^{-2})$:
\begin{multline}
    \left(1+\dfrac{x^2}{2l^2} \right)\Box h_{\alpha\beta} + \dfrac{2}{l^2} h_{\alpha\beta} - \dfrac{x^\gamma\partial_\alpha h_{\beta\gamma}}{l^2} - \dfrac{x^\gamma\partial_\beta h_{\alpha\gamma}}{l^2} + \dfrac{x^\gamma\partial_\gamma h_{\alpha\beta}}{l^2} + \dfrac{x_\beta\partial^\gamma h_{\alpha\gamma}}{l^2} + \dfrac{x_\alpha\partial^\gamma h_{\beta\gamma}}{l^2} = 0,
    \label{Eqn_h}
\end{multline}
where $\Box=\partial^\gamma \partial_\gamma$ is the flat space d'Alembertian operator. The above expression is the flat space linearized equation with perturbative corrections up to $\mathcal{O}(l^{-2})$.

Now, we determine the mode solutions for the linearized field equation by considering the ansatz as follows:
\begin{equation}
\label{f_graviton}
f^h_{\mu\nu}(x,p) = \dfrac{1}{\sqrt{2E_p}} \left(\varepsilon^h_{\mu\nu}(p) + \varepsilon^h_{\mu\alpha} \dfrac{\mathcal{F}^\alpha_\nu(x)}{l^2} + \varepsilon^h_{\alpha\nu} \dfrac{\mathcal{F}_\mu^\alpha(x)}{l^2} + \varepsilon^h_{\alpha\beta}\dfrac{\mathcal{G}^{\alpha\beta}_{\ \ \ \mu\nu}(x)}{l^2} \right) e^{ip\cdot x},
\end{equation}
where $\varepsilon_{\mu\nu}$ is the polarization tensor, $E_p$ is the zeroth component of $p^\mu$, ``$h$" indicates the helicity of the graviton\footnote{The helicity index ``$h$" should not be confused with the trace of $h_{\mu\nu}$.} , and $\mathcal{F}_{\alpha\beta}$ and $\mathcal{G}_{\alpha\beta\mu\nu}$ are arbitrary tensors on the flat space that can be obtained by substituting the ansatz (\ref{f_graviton}) into (\ref{Eqn_h}) and solving it for $\mathcal{F}_{\alpha\beta}$ and $\mathcal{G}_{\alpha\beta\mu\nu}$. We obtained:
\begin{align}
    &\mathcal{F}_{\alpha\beta}(x)= \frac{1}{2}x_\alpha x_\beta - \frac{1}{8} \eta_{\alpha\beta} x^2,\\
    &\mathcal{G}_{\alpha\beta\mu\nu} = -\frac{1}{4}x_\alpha x_\beta \eta_{\mu\nu}, \ \ p^2 = \frac{2}{l^2};
\end{align}
where $\varepsilon^\alpha_\alpha = 0$ and $p^\alpha \varepsilon_{\alpha\beta} = \mathcal{O}(l^{-2})$ have been used to obtain the solution, which are consistent with the transverse-traceless gauge (\ref{TT-gauge}). For $l\to\infty$, the mode solution reduces to the plane wave solution for the flat space linearized field equation.

Let us now check the orthogonality of these modes by defining an inner product on the solution space as follows:
\begin{multline}
\label{graviton_inner_product}
    \left(\mathcal{P}^{\mu\nu\alpha\beta} f^h_{\mu\nu}(p), f^{h'}_{\alpha\beta}(p') \right) = -i \int d^3x \sqrt{-g} \  \mathcal{P}^{\mu\nu\alpha\beta}\ e^{-\epsilon\frac{|\overrightarrow{x}|}{R}} \big(f^{*h}_{\mu\nu}(x,p) \nabla^t f^{h'}_{\alpha\beta}(x,p') \\- f^{h'}_{\mu\nu}(x,p') \nabla^t f^{*h}_{\alpha\beta}(x,p) \big),
\end{multline}
\begin{equation}
    \mathcal{P}^{\mu\nu\alpha\beta} = \dfrac{1}{2}\left(g^{\mu\alpha} g^{\nu\beta} + g^{\mu\beta} g^{\nu\alpha} - g^{\mu\nu}g^{\alpha\beta} \right);
\end{equation}
where the exponential damping factor is introduced to eliminate the boundary terms. Plugging two different modes in the above expression, we have verified that the modes are indeed orthogonal:
\begin{equation}
    \left(\mathcal{P}^{\mu\nu\alpha\beta} f^h_{\mu\nu}(p), f^{h'}_{\alpha\beta}(p') \right) = (2\pi)^3 \delta_{hh'} \delta^{(3)}(p-p').
\end{equation}
Now, we write the mode expansion of the field $h_{\mu\nu}$ as
\begin{equation}
    h_{\mu\nu}(x) = \sum_{h=\pm} \int\dfrac{d^3 p}{(2\pi)^3} \left(a^h_p f^h_{\mu\nu}(x,p) + a^{h\dagger}_p f^{h*}_{\mu\nu}(x,p) \right),
\end{equation}
where $a^h_p$ ($a^{h\dagger}_p$) is the annihilation (creation) operator which acts on the asymptotic states to annihilate (create) a graviton of momentum $p$ and helicity ``$h$".

\subsection{LSZ formula for gravitons}
\label{LSZ_graviton_sec}
In this section, we derive the LSZ reduction formula for gravitons. We can invert the inner product expression to obtain
\begin{equation}
    a_p^h = \left(\mathcal{P}^{\mu\nu\alpha\beta} f^{h}_{\alpha\beta}(x, p), h_{\mu\nu}(x,p') \right).
\end{equation}
Since the scattering processes are confined to a compact region $R$ inside the static patch, the particles can reach the boundaries of $R$ in a finite time $T$ which is much larger than the interaction time scale. Therefore,
\begin{multline*}
    a_p^h(T)-a_p^h(-T) = \int_{-T}^{T} dt\ \partial_t a_p^h = \int_{-T}^{T} dt\ \partial_t \left(\mathcal{P}^{\mu\nu\alpha\beta} f^{h}_{\alpha\beta}(x, p), h_{\mu\nu}(x,p') \right) \\
    = -i\int dt d^3 x\ e^{-\epsilon\frac{|\overrightarrow{x}|}{R}} \partial_t \left(\sqrt{-g} \mathcal{P}^{\mu\nu\alpha\beta}\  \big(f^{*h}_{\alpha\beta}(x,p) \nabla^t h_{\mu\nu}(x,p') - h_{\alpha\beta}(x,p') \nabla^t f^{*h}_{\mu\nu}(x,p) \big) \right) \\
    = -i\int d^4 x\ e^{-\epsilon\frac{|\overrightarrow{x}|}{R}} \partial_\rho \left(\sqrt{-g} \mathcal{P}^{\mu\nu\alpha\beta}\  \big(f^{*h}_{\alpha\beta}(x,p) \nabla^\rho h_{\mu\nu}(x,p') - h_{\alpha\beta}(x,p') \nabla^\rho f^{*h}_{\mu\nu}(x,p) \big) \right) \\
    = -i\int d^4 x e^{-\epsilon\frac{|\overrightarrow{x}|}{R}} \sqrt{-g}\ \nabla_\rho \left( \mathcal{P}^{\mu\nu\alpha\beta}\  \big(f^{*h}_{\alpha\beta}(x,p) \nabla^\rho h_{\mu\nu}(x,p') - h_{\alpha\beta}(x,p') \nabla^\rho f^{*h}_{\mu\nu}(x,p) \big) \right),
\end{multline*}
where in the third line we have added terms at the spatial boundary which are in fact zero. Simplifying the above expression and taking $\epsilon\to 0$, we obtain
\begin{equation}
    a_p^h(T)-a_p^h(-T) = -i\int d^4x \sqrt{-g} \mathcal{P}^{\mu\nu\alpha\beta} f^{*h}_{\mu\nu}(x,p) \left(\nabla^2 - \dfrac{2}{l^2} \right) h_{\alpha\beta},
    \label{g_inner_invert}
\end{equation}
where the operator $\left(\nabla^2 - 2/l^2 \right)$ acts on the graviton propagator (or Green's function) $G_{\mu\nu\alpha\beta}(x-x')$ as follows:
\begin{equation}
    \left(\nabla_x^2 - \dfrac{2}{l^2} \right) G_{\mu\nu\alpha\beta}(x-x') = i\mathcal{P}_{\mu\nu\alpha\beta} \dfrac{\delta^{(4)}(x-x')}{\sqrt{-g(x)}}.
\end{equation}
This relation will be used in section (\ref{soft_graviton_theorem_sec}) to compute the $\mathcal{S}$-matrix.

\section{Perturbative $\mathcal{S}$-matrix and LSZ formula}
\label{S-matrix_sec}
In this section, we define a perturbative $\mathcal{S}$-matrix in the de Sitter space and deduce curved spacetime generalization of the LSZ formula. The proper description of the $\mathcal{S}$-matrix requires well-defined asymptotic states at the boundaries of the spacetime. Since the largest observable region in de Sitter space is the static patch, which is bounded by cosmological horizons, we can not define asymptotic states at the boundaries of the de Sitter space. In such situations, one can treat cosmological horizons as physical boundaries\footnote{In \cite{Teitelboim_2001, Teitelboim_2003}, the cosmological horizons are considered as physical boundaries to study the thermodynamics of a rotating black hole.} of the spacetime to define the asymptotic states. In our analysis, since the scattering processes are confined to the compact region inside the static patch, the asymptotic states are defined at the early- and late-time Cauchy surfaces which act as Hilbert spaces of the incoming and outgoing particles. The particles can reach the Cauchy surface in a finite time $T$, which we assume to be much larger than the interaction time scale. All lengths in our calculations are very small compared to the de Sitter radius $l$, i.e., $|x|<<l$.

The $\mathcal{S}$-matrix describing the scattering of incoming particles with momenta $p_i$ to the outgoing particles with momenta $p_j$ followed by the emission of gravitons with momenta $p_k$ is defined as 
\begin{equation}
\label{Gamma_general}
    \Gamma (\{p_j,p_k\},\{p_i\}) = \lim_{t\to T} \left\langle \mathcal{T} \prod_{j,k\in \text{out}} \sqrt{2E_{p_j}} a_{p_j}(t)\ \sqrt{2E_{p_k}} a_{p_k}(t) \prod_{i\in \text{in}} \sqrt{2E_{p_i}} a^{\dagger}_{p_i}(-t) \right\rangle,
\end{equation}
where $\mathcal{T}$ represents time-ordering.
To deduce the LSZ formula, we rewrite the above expression as follows:
\begin{multline}
\label{Gamma_2}
    \Gamma (\{p_j,p_k\},\{p_i\}) = \lim_{t\to T} \bigg\langle \mathcal{T} \prod_{j,k\in \text{out}} \sqrt{2E_{p_j}} \big(a_{p_j}(t) - a_{p_j}(-t)\big) \sqrt{2E_{p_k}}\big(a_{p_k}(t)-a_{p_k}(-t)\big) \\
    \cdot \prod_{i \in \text{in}} \sqrt{2E_{p_i}} \big(a^{\dagger}_{p_i}(-t)-a^{\dagger}_{p_i}(t)\big) \bigg\rangle,
\end{multline}
where all the annihilation (creation) operators go to the right (left) side in the time order product and annihilate the ``in" (``out") -state, so that the above expression reduces to Eq. (\ref{Gamma_general}). By inverting the inner product expression for the scalar field modes (\ref{scalar_inner_product}), as done for gravitons in section (\ref{LSZ_graviton_sec}), one can deduce \cite{Sayali_Diksha}
\begin{equation}
    a_p(T)-a_p(-T) = -i\int_{-T}^{T} d^4x \sqrt{-g} g_p^*(x) \big(\nabla^2 - m^2 \big)\phi(x).
    \label{s_inner_invert}
\end{equation}
Substituting Eqs. (\ref{s_inner_invert}) and (\ref{g_inner_invert}) into Eq. (\ref{Gamma_2}) yields
\begin{multline}
    \Gamma (\{p_j,p_k\},\{p_i\}) = \int\prod_{j,k\in \text{out}}\prod_{i\in \text{in}} [d^4x_j] [d^4y_k] [d^4z_i]\ g^*_{p_j}(x_j) \mathcal{P}^{\mu\nu\alpha\beta} f^{*}_{\mu\nu}(y_k,p_k) g_{p_i}(z_i)\\
    \cdot(-i) \big(\nabla_{x_j}^2 - m^2\big)\ (-i) \big(\nabla_{y_k}^2 - 2/l^2\big)\ (-i) \big(\nabla_{z_i}^2 - m^2\big)\ \big\langle \mathcal{T} \phi(x_j)h_{\alpha\beta}(y_k)...\phi(z_i) \big\rangle,
    \label{LSZ_curved}
\end{multline}
where the measure is
\begin{equation}
    [d^4x_j] [d^4y_k] [d^4z_i] = d^4x_j d^4y_k d^4z_i \sqrt{(2E_{p_j})(2E_{p_k})(2E_{p_i})}\sqrt{\big(-g(x_j)\big)\big(-g(y_k)\big)\big(-g(z_i)\big)},
\end{equation}
and the helicity index is suppressed. Eq. (\ref{LSZ_curved}) is the curved spacetime generalization of the LSZ formula.

\section{Perturbative corrections to flat space soft graviton theorem}
\label{soft_graviton_theorem_sec}
In this section, we compute perturbative corrections ($\mathcal{O}(l^{-1})$ and $\mathcal{O}(l^{-2})$ corrections) to the flat space soft graviton theorem using the LSZ formula discussed in the previous section, and deduce leading, subleading, and sub-subleading corrections. The action for a massive scalar field minimally coupled to the Einstein gravity is written as
\begin{equation}
S = -\int d^4x \sqrt{-g} \bigg[\frac{2}{\kappa^2}(\mathcal{R} - 2\Lambda)+ \frac{1}{2}g^{\mu\nu} \nabla_\mu\phi \nabla_\nu\phi - \frac{1}{2}m^2\phi^2 + V(\phi) \bigg],
\end{equation}
where $V(\phi)$ is a scalar potential, $g_{\mu\nu}$ is a general spacetime metric, and $\nabla_\mu$ is the covariant derivative compatible with $g_{\mu\nu}$. In the weak field expansion, $g_{\mu\nu} = \Bar{g}_{\mu\nu} + \kappa h_{\mu\nu}$,\footnote{Bar notations are used for the de Sitter space to distinguish it from the general spacetime, which will be dropped later.} the leading order terms are
\begin{align}
    \mathcal{L}_{gravity} &= -\frac{2}{\kappa^2}\sqrt{-g} (\mathcal{R} - 2\Lambda) \nonumber \\
    &= \frac{1}{2}\Bar{\nabla}_\sigma h_{\mu\nu} \Bar{\nabla}^\sigma h^{\mu\nu} - \frac{1}{2}\Bar{\nabla}_\sigma h \Bar{\nabla}^\sigma h - \Bar{\nabla}_\mu h_{\nu\sigma} \Bar{\nabla}^\sigma h^{\mu\nu} + \Bar{\nabla}_\mu h^{\mu\nu} \Bar{\nabla}_\nu h - \frac{3}{l^2} \bigg(h_{\mu\nu}h^{\mu\nu} - \frac{1}{2}h^2 \bigg) + \cdots ,
\end{align}
\begin{align}
    \mathcal{L}_{scalar} &= -\sqrt{-g} \bigg[\frac{1}{2}g^{\mu\nu} \nabla_\mu\phi \nabla_\nu\phi - \frac{1}{2}m^2\phi^2 + V(\phi) \bigg] \nonumber \\
    &= -\frac{1}{2} \Bar{\nabla}_\sigma\phi \Bar{\nabla}^\sigma\phi + \frac{1}{2} m^2\phi^2 - V(\phi) + \frac{\kappa}{2} h^{\mu\nu} \bigg[\Bar{\nabla}_\mu\phi \Bar{\nabla}_\nu\phi - \frac{1}{2} \Bar{g}_{\mu\nu} \big(\Bar{\nabla}_\sigma\phi \Bar{\nabla}^\sigma\phi - m^2\phi^2 \big)  \bigg] + \cdots ,
\end{align}
where the total derivative terms and the constant terms have been dropped in the Lagrangian.

\begin{figure}
    \centering
    \includegraphics[scale=0.25]{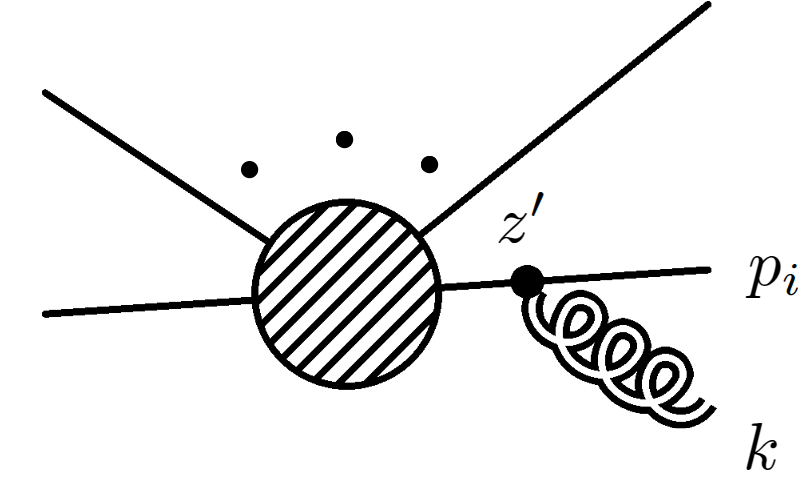}
    \caption{Scattering of scalars with momenta $p_i$ followed by an emission of a soft graviton with momentum $k$.}
    \label{soft_graviton_fig}
\end{figure}

Let us consider a scattering of $n$ scalars with momenta $p_i$ followed by an emission of a soft graviton with momentum $k$, as shown in Fig. (\ref{soft_graviton_fig}). The graviton is attached to an external line by three-point minimal coupling. The graviton attached to an internal line, i.e., Fig. (\ref{soft_internal_fig}), will be analyzed in subsection (\ref{Internal_line_sec}).  We consider all external lines to be outgoing for the ease of the calculations. The $\mathcal{S}$-matrix corresponds to the scattering process depicted in Fig. (\ref{soft_graviton_fig}) is described as follows:\footnote{Hereafter, the bar notations for the de Sitter space have been dropped.}

\begin{multline}
\label{Gamma_LSZ}
\Gamma_{n+1}(\{p_1,...,p_n\},k) = \sum_{i=1}^{n}\frac{\kappa_i}{2} \int d^4z_i \sqrt{-g(z_i)} g^{*}_{p_i}(z_i) (-i) (\nabla^{2}_{z_i}-m^2)\\
\cdot \int d^4y \sqrt{-g(y)} \mathcal{P}^{\mu\nu\alpha\beta} f^{*}_{\mu\nu}(y,k) (-i) (\nabla_y^2 - 2/l^2)\\
\cdot \int\prod_{j=1; j\ne i}^{n-1} d^4z_j \sqrt{-g(z_j)} g^*_{p_j}(z_j)(-i)(\nabla^{2}_{z_j}-m^2) \int d^4z \sqrt{-g(z)} \int d^4z' \sqrt{-g(z')}\\ 
\cdot\bigg\langle \mathcal{T} \phi(z_i) h_{\alpha\beta}(y) h^{\rho\sigma}(z') \bigg[\nabla'_{\rho}\phi(z') \nabla'_{\sigma}\phi(z') - \frac{1}{2}g_{\rho\sigma} \nabla'_{\tau}\phi(z') \nabla'^{\tau}\phi(z') + \frac{1}{2} g_{\rho\sigma} m^2 \phi^2(z') \bigg] G[\phi(z)...\phi(z_j)] \bigg\rangle,
\end{multline}
where $G[\phi(z)...\phi(z_j)]$ is the $n$-point correlation function for scalars which receives contributions from any arbitrary interaction vertices, and $\kappa_i$ is the scalar-graviton coupling constant. The sum is taken over all the scalars which can emit a soft graviton to take into account the contributions from all the external legs. Now, we perform the Wick contractions among the fields and generate the respective propagators as follows:
\begin{multline}
\bigg\langle \mathcal{T} \phi(z_i) h_{\alpha\beta}(y) h^{\rho\sigma}(z') \bigg[\nabla'_{\rho}\phi(z') \nabla'_{\sigma}\phi(z') - \frac{1}{2}g_{\rho\sigma} \nabla'_{\tau}\phi(z') \nabla'^{\tau}\phi(z') + \frac{1}{2} g_{\rho\sigma} m^2 \phi^2(z') \bigg] G[\phi(z)...\phi(z_j)] \bigg\rangle \\
= \bigg\langle G_{\alpha\beta}^{\rho\sigma}(y-z') \big\{2\nabla'_\rho D(z_i,z') \nabla'_\sigma D(z',z) - g_{\rho\sigma} \nabla'_\tau D(z_i,z') \nabla'^\tau D(z',z)\\
+ g_{\rho\sigma} m^2 D(z_i,z') D(z',z) \big\} G[\{\phi(z_j) \}] \bigg\rangle,
\end{multline}
where $G^{\rho\sigma}_{\alpha\beta}(x-y)$ and $D(x,y)$ are graviton and scalar propagators, respectively. Since the propagators are the Green's functions, the equation-of-motion operators in (\ref{Gamma_LSZ}), i.e., $(\nabla_{z_i}^2 - m^2)$ and $(\nabla_y^2 - 2/l^2)$, will act on the respective propagator and produce the delta functions.
\begin{multline}
\Gamma_{n+1}(\{p_1,...,p_n\},k) = \sum_{i=1}^{n}\frac{\kappa_i}{2} \int d^4z_i \sqrt{-g(z_i)} g^{*}_{p_i}(z_i) (-i) \int d^4y \sqrt{-g(y)} \mathcal{P}^{\mu\nu\alpha\beta} f^{*}_{\mu\nu}(y,k) (-i) \\
\cdot \int\prod_{j=1; j\ne i}^{n-1} d^4z_j \sqrt{-g(z_j)} g^*_{p_j}(z_j)(-i)(\nabla^{2}_{z_j}-m^2) \int d^4z \sqrt{-g(z)} \int d^4z' \sqrt{-g(z')}\\ 
\cdot \bigg\langle \mathcal{P}_{\alpha\beta}^{\rho\sigma} \dfrac{i\delta^{(4)}(y-z')}{\sqrt{-g(y)}} \bigg\{2i\partial'_\rho \dfrac{\delta^{(4)}(z_i-z')}{\sqrt{-g(z_i)}} \partial'_\sigma D(z',z) - ig_{\rho\sigma} g^{\tau\lambda}\partial'_\tau \dfrac{\delta^{(4)}(z_i-z')}{\sqrt{-g(z_i)}} \partial'_\lambda D(z',z)\\
+ ig_{\rho\sigma} m^2 \dfrac{\delta^{(4)}(z_i-z')}{\sqrt{-g(z_i)}} D(z',z) \bigg\} G[\{\phi(z_j) \}] \bigg\rangle.
\end{multline}
Let us now evaluate integrations in $y$ and $z_i$.
First, we perform the $y$-integral which is trivial due to the presence of the delta function $\delta^{(4)}(y-z')$. Performing $y$-integral on the delta function will simply replace the coordinate $y$ by $z'$ in the above expression. Similarly, we can perform the $z_i$-integral on another delta function $\delta^{(4)}(z_i-z')$. Since the coordinates $z_i$ and $z'$ are independent, we move the $z_i$-integral inside the derivative operator $\partial'_\alpha$, which acts on the delta function in the above expression, and then perform the integration. We obtain
\begin{multline}
\Gamma_{n+1}(\{p_1,...,p_n\},k) = \int d^4z \sqrt{-g(z)} \int d^4z' \sqrt{-g(z')} \mathcal{P}^{\mu\nu\alpha\beta} \mathcal{P}_{\alpha\beta}^{\rho\sigma} f^{*}_{\mu\nu}(z',k)\\
\cdot \sum_{i=1}^{n}\frac{\kappa_i}{2} \bigg\{2\partial'_\rho g^*_{p_i}(z') \partial'_\sigma D(z',z) - g_{\rho\sigma} g^{\tau\lambda} \partial'_\tau g^*_{p_i}(z') \partial'_\lambda D(z',z)
+ g_{\rho\sigma} m^2 g^*_{p_i}(z') D(z',z) \bigg\} \\
\cdot\int\prod_{j=1; j\ne i}^{n-1} d^4z_j \sqrt{-g(z_j)} g^*_{p_j}(z_j)(-i)(\nabla^{2}_{z_j}-m^2) \big\langle G[\{\phi(z_j) \}] \big\rangle,
\end{multline}
which further simplifies to
\begin{multline}
\label{LSZ_final}
\Gamma_{n+1}(\{p_1,...,p_n\},k) = \sum_{i=1}^{n}\frac{\kappa_i}{2} \int d^4z d^4z' \left(1-\dfrac{z^2}{l^2} - \dfrac{z'^2}{l^2} \right) f^{*\rho\sigma}(z',k) \bigg\{2\partial'_\rho g^*_{p_i}(z') \partial'_\sigma D(z',z) \bigg\} \\
\cdot\int\prod_{j=1; j\ne i}^{n-1} d^4z_j \sqrt{-g(z_j)} g^*_{p_j}(z_j)(-i)(\nabla^{2}_{z_j}-m^2) \big\langle G[\{\phi(z_j) \}] \big\rangle,
\end{multline}
where the traceless condition $g_{\rho\sigma}f^{*\rho\sigma}=0$ is used.

This is the main result from which we compute the leading, subleading, and sub-subleading corrections to the flat space soft factor in the following sections. Substituting all the essential elements, i.e., scalar field modes, graviton modes, and the scalar field propagator, in the above expression, keeping terms up to $\mathcal{O}(l^{-2})$, and taking the soft limit ($k\to 0$ or $\omega\to 0$) produces the final expression that can be put in the following form:
\begin{multline}
\label{Gamma_structure}
\Gamma_{n+1}(\{p_1,...,p_n\},\omega \hat{k}) = \bigg(A^{(0)}(\{p_i\},\omega \hat{k}) + \dfrac{1}{l}A^{(1)}(\{p_i\},\omega \hat{k}) + \dfrac{1}{l^2}A^{(2)}(\{p_i\},\omega \hat{k}) \bigg)\\ \cdot \bigg(\Gamma_n^{(0)}(\{p_1,...,p_n\}) + \dfrac{1}{l}\Gamma_n^{(1)}(\{p_1,...,p_n\}) + \dfrac{1}{l^2} \Gamma_n^{(2)}(\{p_1,...,p_n\}) \bigg),
\end{multline}
where $A^{(1)}$ and $A^{(2)}$ are the perturbative corrections to the flat space soft factor $A^{(0)}$. Similarly, $\Gamma_n^{(1)}$ and $\Gamma_n^{(2)}$ are the perturbative corrections to the flat space $\mathcal{S}$-matrix, $\Gamma_n^{(0)}$, for the scattering of $n$-scalars.

We must note that the perturbative corrections to the flat space soft factor are only valid for $\omega l >> 1$. And the soft limit of scattering amplitudes in flat space requires $\omega<<p_i^\mu$, where $p_i^\mu$ is the momenta of the hard particles. Thus, our analysis, and hence results are valid in the regime $\frac{1}{l}<<\omega<<p_i^\mu$. For the soft graviton scattering in the de Sitter background, we take the double scaling limit \cite{Banerjee_Bhattacharjee_Mitra} in which $\omega\to 0$ and $l\to\infty$ simultaneously, keeping their product $\omega l$ finite. In these limits, it is convenient to introduce a new parameter $\delta=1/\omega l$. To put the perturbative corrections in a proper hierarchy of leading, subleading, and sub-subleading corrections, the full soft factor (i.e. flat space + perturbative corrections) can be expanded in $\delta$ as follows:
\begin{equation}
\label{delta_omega_exp}
    A^{(0)}(\{p_i\},\omega \hat{k}) + \dfrac{1}{l}A^{(1)}(\{p_i\},\omega \hat{k}) + \dfrac{1}{l^2}A^{(2)}(\{p_i\},\omega \hat{k})=\sum_{r=0}^{2} \sum_{s\ge -1} \delta^r S^{(r,s)}(\{p_i\},\omega \hat{k}),
\end{equation}
where `$s$' represents the power of $\omega$, i.e., $S^{(r,s)} \sim \mathcal{O}(\omega^s)$. In this soft expansion, $S^{(0,-1)}$ is the flat space leading soft factor, and the higher order terms are leading, subleading, sub-subleading corrections, and so on.

\subsection{Leading corrections}
\label{leading_sec}
Let us substitute the scalar field propagator (\ref{symmetric_prop}), scalar field modes (\ref{gp_modes}), and graviton modes (\ref{f_graviton}) into Eq. (\ref{LSZ_final}) and simplify the expression by keeping the terms up to $\mathcal{O}(l^{-1})$.
\begin{multline}
\label{Gamma_upto_leading}
\Gamma_{n+1}(\{p_1,...,p_n\},k) = -i\sum_{i=1}^{n}\kappa_i \int d^4z d^4z' \int\dfrac{d^4p}{(2\pi)^4} \bigg[ \dfrac{p^\alpha p_i^\beta \varepsilon_{\alpha\beta}}{p^2+m^2}
+ \dfrac{3i(p^\alpha p_i^\beta \varepsilon_{\alpha\beta})}{2lm(p^2+m^2)} + \dfrac{m(p^\alpha z'^\beta \varepsilon_{\alpha\beta})}{l(p^2+m^2)} \\
+ \dfrac{5(p_i\cdot z')(p^\alpha p_i^\beta \varepsilon_{\alpha\beta})}{2lm(p^2+m^2)}
- \dfrac{i(p_i\cdot z')^2(p^\alpha p_i^\beta \varepsilon_{\alpha\beta})}{2lm(p^2+m^2)}
- \dfrac{im z'^2 (p^\alpha p_i^\beta \varepsilon_{\alpha\beta})}{2l(p^2+m^2)}
\bigg] e^{i(p-p_i-k)\cdot z'} e^{-ip\cdot z} \\
\cdot \int\prod_{j=1; j\ne i}^{n-1} d^4z_j \sqrt{-g(z_j)} g^*_{p_j}(z_j)(-i)(\nabla^{2}_{z_j}-m^2) \big\langle G[\{\phi(z_j) \}] \big\rangle.
\end{multline}
To simplify the above expression, we replace all $z'^\mu$ by $-i\frac{\partial}{\partial p_\mu}$.
\begin{multline}
\label{LSZ_upto_l1}
\Gamma_{n+1}(\{p_1,...,p_n\},k) = -i\sum_{i=1}^{n}\kappa_i \int d^4z d^4z' \int\dfrac{d^4p}{(2\pi)^4} \bigg[ \dfrac{p^\alpha p_i^\beta \varepsilon_{\alpha\beta}}{p^2+m^2}
+ \dfrac{3i(p^\alpha p_i^\beta \varepsilon_{\alpha\beta})}{2lm(p^2+m^2)} - \dfrac{im(p^\alpha \varepsilon_{\alpha\beta})}{l(p^2+m^2)} \dfrac{\partial}{\partial p_\beta} \\
- \dfrac{5i(p^\alpha p_i^\beta \varepsilon_{\alpha\beta})}{2lm(p^2+m^2)} p_i^\gamma \dfrac{\partial}{\partial p^\gamma}
+ \dfrac{i(p^\alpha p_i^\beta \varepsilon_{\alpha\beta})}{2lm(p^2+m^2)}p_i^\gamma p_i^\delta \dfrac{\partial}{\partial p^\gamma} \dfrac{\partial}{\partial p^\delta}
+ \dfrac{im (p^\alpha p_i^\beta \varepsilon_{\alpha\beta})}{2l(p^2+m^2)} \dfrac{\partial}{\partial p^\gamma} \dfrac{\partial}{\partial p_\gamma} \bigg] e^{i(p-p_i-k)\cdot z'} \\
\cdot e^{-ip\cdot z}
\int\prod_{j=1; j\ne i}^{n-1} d^4z_j \sqrt{-g(z_j)} g^*_{p_j}(z_j)(-i)(\nabla^{2}_{z_j}-m^2) \big\langle G[\{\phi(z_j) \}] \big\rangle.
\end{multline}
The above expression contains $\mathcal{O}(l^0)$ and $\mathcal{O}(l^{-1})$ terms. We analyze them step by step. Let us first evaluate the $\mathcal{O}(l^0)$ term, which should produce the flat space soft factor.
\begin{multline}
\Gamma^{(0)}_{n+1}(\{p_1,...,p_n\},k) = -i\sum_{i=1}^{n}\kappa_i \int d^4z d^4z' \int\dfrac{d^4p}{(2\pi)^4} \dfrac{p^\alpha p_i^\beta \varepsilon_{\alpha\beta}}{p^2+m^2} e^{i(p-p_i-k)\cdot z'} e^{-ip\cdot z}\\
\cdot\int\prod_{j=1; j\ne i}^{n-1} d^4z_j e^{-ip_j\cdot z_j} (-i)(\nabla^{2}_{z_j}-m^2) \big\langle G[\{\phi(z_j) \}] \big\rangle.
\end{multline}
This expression is trivial to simplify. By performing the $z'$-integral in the above expression, one can generate the delta function $\delta^{(4)}(p-p_i-k)$, which can be integrated w.r.t. $p$ to replace all $p$ by $p_i+k$ in the resultant expression. Finally, by taking the soft limit, one can obtain
\begin{align}
\Gamma^{(0)}_{n+1}(\{p_1,...,p_n\}, k) = &-i \sum_{i=1}^{n}\frac{\kappa_i}{2} \dfrac{p_i^\alpha p_i^\beta\varepsilon_{\alpha\beta}}{p_i\cdot k} \int d^4z e^{-ip_i\cdot z}\nonumber \\
&\cdot\int\prod_{j=1; j\ne i}^{n-1} d^4z_j e^{-i p_j\cdot z_j}(-i)(\nabla^{2}_{z_j}-m^2) \big\langle G[\{\phi(z_j) \}] \big\rangle.
\end{align}
The momentum $p_i$ in the exponential is on-shell, and hence it will remain the same for all the scalars. Thus, we can move $z$-integral outside the summation and rewrite the above expression as
\begin{equation}
\Gamma^{(0)}_{n+1}(\{p_1,...,p_n\}, k) = \sum_{i=1}^{n}\frac{\kappa_i}{2} \dfrac{p_i^\alpha p_i^\beta\varepsilon_{\alpha\beta}}{p_i\cdot k} \Gamma^{(0)}_n{(\{p_1,...,p_n\})},
\end{equation}
which is the flat space soft graviton theorem. One can apply the Ward identity to the above expression and deduce the equivalence principle, i.e., gravity equally couples to all the matter. Thus, $\kappa_1=...=\kappa_n=\kappa$. \\
Therefore, the flat space soft factor is
\begin{equation}
\label{A0_soft_factor}
    A^{(0)} = \sum_{i=1}^{n}\frac{\kappa}{2} \dfrac{p_i^\alpha p_i^\beta\varepsilon_{\alpha\beta}}{p_i\cdot k},
\end{equation}
and from (\ref{delta_omega_exp}), we can write
\begin{equation}
    S^{(0,-1)} = \sum_{i=1}^{n}\frac{\kappa}{2} \dfrac{p_i^\alpha p_i^\beta\varepsilon_{\alpha\beta}}{\omega p_i\cdot \hat{k}}.
\end{equation}

Let us now evaluate the $\mathcal{O}(l^{-1})$ terms in (\ref{LSZ_upto_l1}).
The second term in (\ref{LSZ_upto_l1}) is trivial like the first term and can be evaluated similarly. However, the third and all the subsequent terms are non-trivial and require integration by parts in $p$ before performing the $z'$-integral.
We can not simply perform the $z'$-integral to generate the delta function, because the momentum derivative operator is acting on the exponential. Therefore, we first do the integration by parts in $p$ to move the exponential outside the derivative operator, and then perform the $z'$-integral. For instance, the third term in (\ref{LSZ_upto_l1}) is evaluated as follows:
\begin{align}
    \int d^4z d^4z' &\int\dfrac{d^4p}{(2\pi)^4} \dfrac{p^\alpha \varepsilon_{\alpha\beta}}{p^2+m^2} e^{-ip\cdot z} \left(-\dfrac{im}{l} \dfrac{\partial}{\partial p_\beta} \right)e^{i(p-p_i-k)\cdot z'} \nonumber \\
    &=\dfrac{im}{l} \int d^4z \int\dfrac{d^4p}{(2\pi)^4}\  \dfrac{\partial}{\partial p_\beta} \left(\dfrac{p^\alpha \varepsilon_{\alpha\beta}}{p^2+m^2} e^{-ip\cdot z} \right)\int d^4z' e^{i(p-p_i-k)\cdot z'} \nonumber \\
    &=\dfrac{im}{l} \int d^4z \int\dfrac{d^4p}{(2\pi)^4}\  \dfrac{\partial}{\partial p_\beta} \left(\dfrac{p^\alpha \varepsilon_{\alpha\beta}}{p^2+m^2} e^{-ip\cdot z} \right) (2\pi)^4 \delta^{(4)}(p-p_i-k) \nonumber \\
    &=\dfrac{im}{l} \int d^4z \left(- i\dfrac{p_i^\alpha z^{\beta} \varepsilon_{\alpha\beta}}{2p_i\cdot k} -\dfrac{p_i^\alpha p_i^\beta \varepsilon_{\alpha\beta}}{2(p_i\cdot k)^2} \right) e^{-ip_i\cdot z},
\end{align}
where in the last line, the $p$-integral is performed and the soft limit is applied. We similarly evaluate all the remaining terms in (\ref{LSZ_upto_l1}). The only difference is that for the last two terms, we have to do integration by parts twice to put the exponential factor outside the derivative operators. After evaluating and collecting all the terms in (\ref{LSZ_upto_l1}), we obtain
\vspace{0cm}
\begin{multline}
\label{Gamma_upto_leading_final}
\Gamma_{n+1}(\{p_1,...,p_n\},k) = -i\sum_{i=1}^{n}\frac{\kappa}{2} \int d^4z \bigg(\dfrac{p_i^\alpha p_i^\beta\varepsilon_{\alpha\beta}}{p_i\cdot k} - \dfrac{m}{l} \dfrac{(p_i^\alpha p_i^\beta \varepsilon_{\alpha\beta})(k\cdot z)}{(p_i\cdot k)^2} + \dfrac{2m}{l} \dfrac{p_i^\alpha z^\beta\varepsilon_{\alpha\beta}}{p_i\cdot k} \bigg) e^{-ip_i\cdot z}\\
\cdot\int\prod_{j=1; j\ne i}^{n-1} d^4z_j e^{-ip_j\cdot z_j} (-i)(\nabla^{2}_{z_j}-m^2) \big\langle G[\{\phi(z_j) \}] \big\rangle + \dfrac{1}{l} A^{(0)} \Gamma^{(1)}_n(\{p_1,...,p_n\})\\
= \sum_{i=1}^n \frac{\kappa}{2} \bigg(\dfrac{p_i^\alpha p_i^\beta\varepsilon_{\alpha\beta}}{p_i\cdot k} - \dfrac{im}{l} \dfrac{p_i^\alpha p_i^\beta\varepsilon_{\alpha\beta}}{(p_i\cdot k)^2} k\cdot\partial_{p_i} + \dfrac{2mi}{l} \dfrac{p_i^\alpha \varepsilon_{\alpha\beta}}{p_i\cdot k}\partial_{p_i}^\beta \bigg)\Gamma^{(0)}_n(\{p_1,...,p_n\}) + \dfrac{1}{l} A^{(0)} \Gamma^{(1)}_n(\{p_1,...,p_n\}).
\end{multline}
Where, in the third line, all $z$ coordinates are replaced by $i\partial_{p_i}$ to transform the soft factor in the momentum space. The $\mathcal{O}(l^{-1})$ correction to the flat space soft factor is
\begin{equation}
    A^{(1)} = -im\sum_{i=1}^n \frac{\kappa}{2} \bigg(\dfrac{p_i^\alpha p_i^\beta\varepsilon_{\alpha\beta}}{(p_i\cdot k)^2} k\cdot\partial_{p_i} - \dfrac{2p_i^\alpha \varepsilon_{\alpha\beta}}{p_i\cdot k}\partial_{p_i}^\beta \bigg),
\end{equation}
and from (\ref{delta_omega_exp}), the leading correction to the flat space soft factor is
\begin{equation}
    S^{(1,0)} = -im\sum_{i=1}^n \frac{\kappa}{2} \bigg(\dfrac{p_i^\alpha p_i^\beta\varepsilon_{\alpha\beta}}{(p_i\cdot \hat{k})^2} \hat{k}\cdot\partial_{p_i} - \dfrac{2p_i^\alpha \varepsilon_{\alpha\beta}}{p_i\cdot \hat{k}}\partial_{p_i}^\beta \bigg).
\end{equation}
In subsection (\ref{Internal_line_sec}), we will see that the leading corrections receive contributions from the diagram where soft graviton is attached to an internal line, i.e., Fig. (\ref{soft_internal_fig}).

\subsection{Subleading and sub-subleading corrections}
\label{subleading_sec}
In this section, we determine the subleading and sub-subleading corrections to the flat space soft graviton theorem. We use the asymmetric form of the scalar field propagator (\ref{asymmetric_prop}) for the ease of the calculations. Let us substitute the soft graviton modes (\ref{f_graviton}), scalar field modes (\ref{gp_modes}), and the scalar field propagator (\ref{asymmetric_prop}) into (\ref{LSZ_final}) and separate the $\mathcal{O}(l^{-2})$ terms as follows:
\begin{align}
\label{Gamma_subleading}
&\Gamma_{n+1}^{(2)}(\{p_1,...,p_n\},k) = -i\sum_{i=1}^{n}\kappa \int d^4z d^4z' \int\dfrac{d^4p}{(2\pi)^4} \bigg(
\dfrac{4m^4(p^\alpha p_i^\beta \varepsilon_{\alpha\beta})}{(p^2+m^2)^4} + \dfrac{2m^2(p^\alpha p_i^\beta \varepsilon_{\alpha\beta})}{(p^2+m^2)^3} \nonumber \\
&+  \dfrac{2(p^\alpha p_i^\beta \varepsilon_{\alpha\beta})}{(p^2+m^2)^2} - \dfrac{17p^\alpha p_i^\beta \varepsilon_{\alpha\beta}}{8m^2(p^2+m^2)} + \dfrac{39i(p_i\cdot z')(p^\alpha p_i^\beta \varepsilon_{\alpha\beta})}{8m^2(p^2+m^2)} +  \dfrac{41(p_i\cdot z')^2(p^\alpha p_i^\beta \varepsilon_{\alpha\beta})}{8m^2(p^2+m^2)} \nonumber \\
&-\dfrac{19i(p_i\cdot z')^3(p^\alpha p_i^\beta \varepsilon_{\alpha\beta})}{12m^2(p^2+m^2)}
- \dfrac{(p_i\cdot z')^4(p^\alpha p_i^\beta \varepsilon_{\alpha\beta})}{8m^2(p^2+m^2)} +\dfrac{3z'^2(p^\alpha p_i^\beta \varepsilon_{\alpha\beta})}{2(p^2+m^2)} - \dfrac{3iz'^2(p_i\cdot z')(p^\alpha p_i^\beta \varepsilon_{\alpha\beta})}{2(p^2+m^2)} \nonumber \\
&- \dfrac{z'^2(p_i\cdot z')^2(p^\alpha p_i^\beta \varepsilon_{\alpha\beta})}{4(p^2+m^2)} - \dfrac{m^2z'^4(p^\alpha p_i^\beta \varepsilon_{\alpha\beta})}{8(p^2+m^2)}
- \dfrac{2im^2(p\cdot z)(p^\alpha p_i^\beta \varepsilon_{\alpha\beta})}{(p^2+m^2)^3}
- \dfrac{i(p\cdot z)(p^\alpha p_i^\beta \varepsilon_{\alpha\beta})}{(p^2+m^2)^2} \nonumber \\
&+ \dfrac{m^2z^2(p^\alpha p_i^\beta \varepsilon_{\alpha\beta})}{2(p^2+m^2)^2}
- \dfrac{z^2(p^\alpha p_i^\beta \varepsilon_{\alpha\beta})}{2(p^2+m^2)}
+ \dfrac{3i(p^\alpha z'^\beta \varepsilon_{\alpha\beta})}{2(p^2+m^2)}
+ \dfrac{5(p_i\cdot z')(p^\alpha z'^\beta \varepsilon_{\alpha\beta})}{2(p^2+m^2)}
- \dfrac{i(p_i\cdot z')^2(p^\alpha z'^\beta \varepsilon_{\alpha\beta})}{2(p^2+m^2)} \nonumber \\
&- \dfrac{im^2z'^2(p^\alpha z'^\beta \varepsilon_{\alpha\beta})}{2(p^2+m^2)}
+ \dfrac{(p\cdot z')(p_i^\alpha z'^\beta \varepsilon_{\alpha\beta})}{2(p^2+m^2)}
- \dfrac{(p\cdot p_i) (z'^\alpha z'^\beta \varepsilon_{\alpha\beta})}{4(p^2+m^2)} \bigg) e^{i(p-p_i-k)\cdot z'} \nonumber \\
&\cdot e^{-ip\cdot z} \int\prod_{j=1; j\ne i}^{n-1} d^4z_j \sqrt{-g(z_j)} g^*_{p_j}(z_j)(-i)(\nabla^{2}_{z_j}-m^2) \big\langle G[\{\phi(z_j) \}] \big\rangle.
\end{align}
The terms inside the parentheses in the above expression can be simplified in a similar manner as we did for $\mathcal{O}(l^{-1})$ terms in (\ref{Gamma_upto_leading}) in the previous section. In brief, we first replace all $z'$ by $-i\partial_{p}$, and then do integration by parts in $p$ to move exponential factor $e^{ip\cdot z'}$ outside the derivative operator $\partial_p$. We then combine all the exponential factors to $e^{i(p-p_i-k)\cdot z'}$ and perform the $z'$-integral to generate the delta function $\delta^{(4)}(p-p_i-k)$. Finally, we perform $p$-integral on the delta function and take the soft limit.

We evaluated each term inside the parentheses in (\ref{Gamma_subleading}) in a similar manner as described above and obtained the following expression:
\begin{align}
\label{Gamma_subleading_soft}
&\Gamma_{n+1}^{(2)}(\{p_1,...,p_n\},k) = -i\sum_{i=1}^{n} \frac{\kappa}{2} \int d^4z \bigg(
\dfrac{3m^2(p_i^\alpha p_i^\beta\varepsilon_{\alpha\beta})}{2(p_i\cdot k)^3}
+\dfrac{p_i^\alpha p_i^\beta\varepsilon_{\alpha\beta}}{2(p_i\cdot k)^2}
-\dfrac{17(p_i^\alpha p_i^\beta\varepsilon_{\alpha\beta})}{8m^2(p_i\cdot k)} \nonumber \\
&+ \dfrac{im^2(p_i^\alpha p_i^\beta\varepsilon_{\alpha\beta})(k\cdot z)}{2(p_i\cdot k)^3}
- \dfrac{2i(p_i^\alpha p_i^\beta\varepsilon_{\alpha\beta})(k\cdot z)}{(p_i\cdot k)^2}
+ \dfrac{m^2(p_i^\alpha p_i^\beta\varepsilon_{\alpha\beta})(k\cdot z)^2}{(p_i\cdot k)^3}
- \dfrac{3i(p_i^\alpha p_i^\beta \varepsilon_{\alpha\beta})(p_i\cdot z)}{2(p_i\cdot k)^2} \nonumber \\
&+ \dfrac{39i(p_i^\alpha p_i^\beta \varepsilon_{\alpha\beta})(p_i\cdot z)}{8m^2(p_i\cdot k)}
- \dfrac{2(p_i^\alpha p_i^\beta \varepsilon_{\alpha\beta})(p_i\cdot z)(k\cdot z)}{(p_i\cdot k)^2} 
+ \dfrac{41(p_i^\alpha p_i^\beta\varepsilon_{\alpha\beta})(p_i\cdot z)^2}{8m^2(p_i\cdot k)} 
+ \dfrac{i(p_i^\alpha p_i^\beta\varepsilon_{\alpha\beta})(p_i\cdot z)^2(k\cdot z)}{2(p_i\cdot k)^2} \nonumber \\
&- \dfrac{19i(p_i^\alpha p_i^\beta\varepsilon_{\alpha\beta})(p_i\cdot z)^3}{12m^2(p_i\cdot k)}
- \dfrac{(p_i^\alpha p_i^\beta\varepsilon_{\alpha\beta})(p_i\cdot z)^4}{8m^2(p_i\cdot k)} 
+ \dfrac{z^2 (p_i^\alpha p_i^\beta \varepsilon_{\alpha\beta})}{p_i\cdot k}
+ \dfrac{im^2 z^2 (p_i^\alpha p_i^\beta \varepsilon_{\alpha\beta})(k\cdot z)}{2(p_i\cdot k)^2} \nonumber \\
&- \dfrac{3i z^2 (p_i^\alpha p_i^\beta \varepsilon_{\alpha\beta})(p_i\cdot z)}{2(p_i\cdot k)}
- \dfrac{z^2 (p_i^\alpha p_i^\beta \varepsilon_{\alpha\beta})(p_i\cdot z)^2}{4(p_i\cdot k)} 
- \dfrac{m^2 z^4 (p_i^\alpha p_i^\beta \varepsilon_{\alpha\beta})}{8(p_i\cdot k)} 
- \dfrac{im^2(p_i^\alpha z^\beta\varepsilon_{\alpha\beta})}{2(p_i\cdot k)^2}
+ \dfrac{4i(p_i^\alpha z^\beta\varepsilon_{\alpha\beta})}{p_i\cdot k} \nonumber \\
&- \dfrac{2m^2(p_i^\alpha z^\beta\varepsilon_{\alpha\beta})(k\cdot z)}{(p_i\cdot k)^2}
+ \dfrac{(p_i^\alpha z^\beta\varepsilon_{\alpha\beta})(k\cdot z)}{2(p_i\cdot k)} 
+ \dfrac{5(p_i^\alpha z^\beta\varepsilon_{\alpha\beta})(p_i\cdot z)}{p_i\cdot k}
- \dfrac{i(p_i^\alpha z^\beta\varepsilon_{\alpha\beta})(p_i\cdot z)^2}{p_i\cdot k} \nonumber \\
&- \dfrac{im^2 z^2 (p_i^\alpha z^\beta\varepsilon_{\alpha\beta})}{p_i\cdot k}
+ \dfrac{5m^2(z^\alpha z^\beta \varepsilon_{\alpha\beta})}{4(p_i\cdot k)}
- \dfrac{z^\alpha z^\beta \varepsilon_{\alpha\beta}}{4}
\bigg) \bigg(1-ik\cdot z -\dfrac{1}{2} (k\cdot z)^2 + ... \bigg) e^{-ip_i \cdot z} \nonumber \\
&\cdot \int\prod_{j=1; j\ne i}^{n-1} d^4z_j \sqrt{-g(z_j)} g^*_{p_j}(z_j)(-i)(\nabla^{2}_{z_j}-m^2) \big\langle G[\{\phi(z_j) \}] \big\rangle,
\end{align}
which can be rewritten in the desired form as\footnote{
We must note that from (\ref{Gamma_structure}), the structure of $\Gamma_{n+1}^{(2)}$ is as follows: $\Gamma_{n+1}^{(2)}=A^{(2)}\Gamma_{n}^{(0)}+A^{(1)}\Gamma_{n}^{(1)}+A^{(0)}\Gamma_{n}^{(2)}$. To put Eq. (\ref{Gamma_subleading_soft}) in the above desired form, one needs to identify the terms corresponding to $A^{(0)}$ and $A^{(1)}$ in (\ref{Gamma_subleading_soft}) and separate them from the remaining terms, which correspond to $A^{(2)}$, to reach at the final expression (\ref{Gamma_subleading_soft_final}).}

\begin{align}
\label{Gamma_subleading_soft_final}
&\Gamma_{n+1}^{(2)}(\{p_1,...,p_n\},k) = -i\sum_{i=1}^{n} \frac{\kappa}{2} \int d^4z \bigg(
\dfrac{3m^2(p_i^\alpha p_i^\beta\varepsilon_{\alpha\beta})}{2(p_i\cdot k)^3}
+\dfrac{p_i^\alpha p_i^\beta\varepsilon_{\alpha\beta}}{2(p_i\cdot k)^2} 
- \dfrac{im^2(p_i^\alpha p_i^\beta\varepsilon_{\alpha\beta})(k\cdot z)}{(p_i\cdot k)^3} \nonumber \\
&- \dfrac{3i(p_i^\alpha p_i^\beta \varepsilon_{\alpha\beta})(p_i\cdot z)}{2(p_i\cdot k)^2}
- \dfrac{im^2(p_i^\alpha z^\beta\varepsilon_{\alpha\beta})}{2(p_i\cdot k)^2}
+ \mathcal{O}(k^{-1}) \bigg) e^{-ip_i \cdot z} \nonumber \\
&\cdot\int\prod_{j=1; j\ne i}^{n-1} d^4z_j e^{-ip_j\cdot z_j}(-i)(\nabla^{2}_{z_j}-m^2) \big\langle G[\{\phi(z_j) \}] \big\rangle
+ A^{(1)} \Gamma_n^{(1)} + A^{(0)} \Gamma_n^{(2)},
\end{align}
where $\mathcal{O}(k^{-1})$ terms contribute beyond the sub-subleading order and are suppressed.
To transform the soft factor in the momentum space, we replace all $z$ by $i\partial_{p_i}$ in the above expression. Thus, the $\mathcal{O}(l^{-2})$ correction to the flat space soft factor takes the following form:
\begin{align}
A^{(2)} = \sum_{i=1}^{n} \frac{\kappa}{2} \bigg(
&\dfrac{3m^2(p_i^\alpha p_i^\beta\varepsilon_{\alpha\beta})}{2(p_i\cdot k)^3} + \dfrac{p_i^\alpha p_i^\beta\varepsilon_{\alpha\beta}}{2(p_i\cdot k)^2} + \dfrac{m^2(p_i^\alpha p_i^\beta \varepsilon_{\alpha\beta})}{(p_i\cdot k)^3} k\cdot\partial_{p_i} + \dfrac{3(p_i^\alpha p_i^\beta\varepsilon_{\alpha\beta})}{2(p_i\cdot k)^2}p_i\cdot \partial_{p_i} \nonumber \\
&+ \dfrac{m^2(p_i^\alpha \varepsilon_{\alpha\beta})}{2(p_i\cdot k)^2}\partial_{p_i}^\beta \bigg) + \mathcal{O}(k^{-1}).
\end{align}
We note that $A^{(2)}$ includes two types of terms: $1/k^3$ and $1/k^2$, which obviously do not contribute at the same order. Thus, as in (\ref{delta_omega_exp}), we expand the soft factor in $\delta$ to separate the subleading and sub-subleading order corrections to the flat space soft factor in the above expression. The subleading and sub-subleading corrections take the following forms, respectively.
\begin{equation}
\label{S^{(2,-1)}}
S^{(2,-1)} = \sum_{i=1}^{n} \frac{\kappa}{2} \dfrac{3m^2(p_i^\alpha p_i^\beta \varepsilon_{\alpha\beta})}{2\omega(p_i\cdot \hat{k})^3},
\end{equation}
\begin{equation}
\label{S^{(2,0)}}
S^{(2,0)} = \sum_{i=1}^{n} \frac{\kappa}{2} \bigg(
\dfrac{p_i^\alpha p_i^\beta\varepsilon_{\alpha\beta}}{2(p_i\cdot \hat{k})^2} + \dfrac{m^2(p_i^\alpha p_i^\beta \varepsilon_{\alpha\beta})}{(p_i\cdot \hat{k})^3} \hat{k}\cdot\partial_{p_i} + \dfrac{3(p_i^\alpha p_i^\beta\varepsilon_{\alpha\beta})}{2(p_i\cdot \hat{k})^2}p_i\cdot \partial_{p_i} + \dfrac{m^2(p_i^\alpha \varepsilon_{\alpha\beta})}{2(p_i\cdot \hat{k})^2}\partial_{p_i}^\beta \bigg).
\end{equation}
In the following section, we evaluate the Feynman diagram where a soft graviton is attached to an internal line.\\

\subsection{Graviton attached to an internal line}
\label{Internal_line_sec}
\begin{figure}
    \centering
    \includegraphics[scale=0.17]{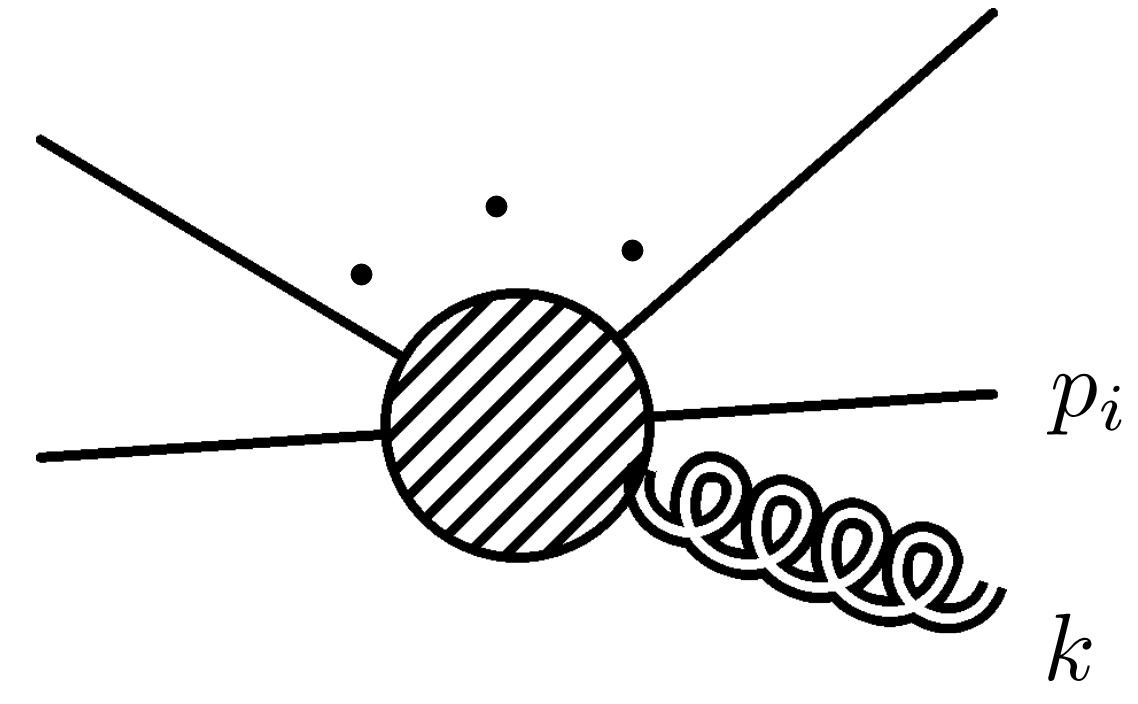}
    \caption{Soft graviton attached to an internal line.}
    \label{soft_internal_fig}
\end{figure}
Let us now consider the Feynman diagram in which the graviton is attached to an internal line (see Fig. (\ref{soft_internal_fig})), which produces the subleading soft graviton theorem in the flat space. We need to check whether such a diagram will introduce additional perturbative corrections to the flat space soft factor in de Sitter background. If the $\mathcal{S}$-matrix for this scattering process is defined by $N_{n+1}(\{p_1,...,p_n\},k)$, then it can be decomposed in the soft expansion as follows:
\begin{multline}
    N_{n+1}(\{p_1,...,p_n\},k) = 
    N^{(0)}_{n+1}(\{p_1,...,p_n\},\omega \hat{k}) + \delta N^{(1)}_{n+1}(\{p_1,...,p_n\}, \omega\hat{k})\\
    + \delta^2 N^{(2)}_{n+1}(\{p_1,...,p_n\},\omega \hat{k}).
\end{multline}
Since the soft graviton is not attached to an external line, this diagram does not produce $\frac{1}{\omega}$-pole terms. Thus, the above expression can be expanded at $\omega\to 0$ as follows:
\begin{multline}
    N_{n+1}(\{p_1,...,p_n\},k) = N^{(0)}_{n+1}(\{p_1,...,p_n\}, 0) + \delta N^{(1)}_{n+1}(\{p_1,...,p_n\}, 0)\\
    + \delta^2 N^{(2)}_{n+1}(\{p_1,...,p_n\}, 0) + ...,
\end{multline}
where $N^{(0)}_{n+1}$ produce the flat space subleading soft factor, $N^{(1)}_{n+1}$ and $N^{(2)}_{n+1}$ are correction terms. Thus, the total $\mathcal{S}$-matrix, after including these additional terms to Eq. (\ref{LSZ_final}), is
\begin{multline}
\Gamma_{n+1}(\{p_1,...,p_n\},k) = \sum_{i=1}^{n}\frac{\kappa_i}{2} \int d^4z d^4z' \left(1-\dfrac{z^2}{l^2} - \dfrac{z'^2}{l^2} \right) f^{*\rho\sigma}(z',k) \bigg\{2\partial'_\rho g^*_{p_i}(z') \partial'_\sigma D(z',z) \bigg\} \\
\cdot\int\prod_{j=1; j\ne i}^{n-1} d^4z_j \sqrt{-g(z_j)} g^*_{p_j}(z_j)(-i)(\nabla^{2}_{z_j}-m^2) \big\langle G[\{\phi(z_j) \}] \big\rangle + N_{n+1}(\{p_1,...,p_n\},k).
\end{multline}
The $\mathcal{S}$-matrix for the scattering process depicted in Fig. (\ref{soft_graviton_fig}) has already been computed in Section (\ref{leading_sec}). Thus, we will not re-do these calculations, and simply add the additional correction terms to Eq. (\ref{Gamma_upto_leading_final}). The resultant expression is
\begin{multline}
\label{LSZ_final_internal}
\Gamma_{n+1}(\{p_1,...,p_n\},k) = \sum_{i=1}^n \frac{\kappa}{2} \bigg(\dfrac{p_i^\alpha p_i^\beta\varepsilon_{\alpha\beta}}{p_i\cdot k} + \dfrac{p_i^\alpha p_i^\beta\varepsilon_{\alpha\beta}}{p_i\cdot k} k\cdot\partial_{p_i} - \dfrac{im}{l} \dfrac{p_i^\alpha p_i^\beta\varepsilon_{\alpha\beta}}{(p_i\cdot k)^2} k\cdot\partial_{p_i} + \dfrac{2mi}{l} \dfrac{p_i^\alpha \varepsilon_{\alpha\beta}}{p_i\cdot k}\partial_{p_i}^\beta \bigg) \Gamma^{(0)}_n\\
+  \dfrac{1}{l} A^{(0)} \Gamma^{(1)}_n + \varepsilon_{\alpha\beta} N^{(0)\alpha\beta}_{n+1}(\{p_1,...,p_n\}, 0) + \delta \varepsilon_{\alpha\beta} N^{(1)\alpha\beta}_{n+1}(\{p_1,...,p_n\}, 0),
\end{multline}
where the second term arises from the soft expansion of the exponential factor $e^{-ik\cdot z'}$, and $N_{n+1}$ is decomposed as $\varepsilon_{\alpha\beta}N_{n+1}^{\alpha\beta}$.
The gauge invariance allows us to evaluate additional correction terms $N^{(0)\alpha\beta}_{n+1}$ and $N^{(1)\alpha\beta}_{n+1}$. The gauge transformation does not receive perturbative corrections at this order, i.e.,  $\varepsilon_{\alpha\beta} \to \varepsilon_{\alpha\beta} - ik_\alpha\xi_\beta - ik_\beta\xi_\alpha$. Thus, under this gauge transformation, the $\mathcal{S}$-matrix should remain invariant, i.e., 
\begin{equation}
\label{gauge_invariance_S-matrix}
    \sum_{i=1}^n \frac{\kappa}{2} \bigg(2(p_i\cdot \xi) + 2(p_i\cdot \xi) k\cdot\partial_{p_i} + \dfrac{2mi}{l} \xi_\beta \partial^\beta_{p_i} \bigg) \Gamma^{(0)}_n
    + 2 k_\alpha \xi_\beta N^{(0)\alpha\beta}_{n+1} + \dfrac{2}{\omega l} k_\alpha \xi_\beta N^{(1)\alpha\beta}_{n+1} = 0,
\end{equation}
where the first term vanishes due to the momentum conservation. Equating the remaining terms order by order, we obtain\footnote{After deducing $N_{n+1}^{(0)\alpha\beta}$ and $N_{n+1}^{(1)\alpha\beta}$ from Eq. (\ref{gauge_invariance_S-matrix}), we have put them in the symmetrized form in Eqs. (\ref{N0}) and (\ref{N1}).}
\begin{align}
    &N_{n+1}^{(0)\alpha\beta} = -\dfrac{1}{2}\sum_{i=1}^{n} \dfrac{\kappa}{2} \big(p_i^\alpha \partial^{\beta}_{p_i} + p_i^\beta \partial^{\alpha}_{p_i} \big) \Gamma_n^{(0)}, \label{N0} \\
    &N_{n+1}^{(1)\alpha\beta} = -\dfrac{im}{2}\sum_{i=1}^{n} \dfrac{\kappa}{2} \bigg(\dfrac{p_i^\alpha}{p_i\cdot \hat{k}} \partial^\beta_{p_i} + \dfrac{p_i^\beta}{p_i\cdot \hat{k}} \partial^\alpha_{p_i} \bigg)\Gamma_n^{(0)}. \label{N1}
\end{align}
These are the additional contributions to the flat space soft graviton theorem from the diagram where soft graviton is attached to an internal line. After plugging these corrections, Eqs. (\ref{N0}) and (\ref{N1}), into Eq. (\ref{LSZ_final_internal}), we obtained the final expression:
\begin{equation}
\Gamma_{n+1}(\{p_1,...,p_n\},k) = \sum_{i=1}^n \frac{\kappa}{2} \bigg(
\dfrac{p_i^\alpha p_i^\beta\varepsilon_{\alpha\beta}}{p_i\cdot k} 
-i \dfrac{p_i^\alpha\varepsilon_{\alpha\beta}k_\gamma J^{\beta\gamma}}{p_i\cdot k} 
- \dfrac{m}{l} \dfrac{p_i^\alpha\varepsilon_{\alpha\beta}k_\gamma J^{\beta\gamma}}{(p_i\cdot k)^2} \bigg) \Gamma^{(0)}_n
+  \dfrac{1}{l} A^{(0)} \Gamma^{(1)}_n,
\end{equation}
where $J^{\beta\gamma}=i\big(p_i^{\beta}\partial^\gamma_{p_i} - p_i^{\gamma}\partial^\beta_{p_i} \big)$ is the angular momentum operator in the momentum space. The second term in the above expression is the flat space subleading soft graviton theorem, and the third term is the leading correction to the flat space soft factor.
\begin{equation}
\label{S^{(1,0)}_final}
S^{(1,0)} = -m \sum_{i=1}^n \frac{\kappa}{2} \dfrac{p_i^\alpha\varepsilon_{\alpha\beta}\hat{k}_\gamma J^{\beta\gamma}}{(p_i\cdot \hat{k})^2}
\end{equation}
We must note that the subleading corrections to the flat space soft factor do not receive additional contributions from the diagram where soft graviton is attached to an internal line. It is because the $N_{n+1}^{(2)}$ correction term arises at order $\mathcal{O}(\delta^2)$ in the soft expansion. Thus, the subleading corrections (\ref{S^{(2,-1)}}) remain the same even after including the contributions from the diagram in Fig. (\ref{soft_internal_fig}). To check whether the sub-subleading corrections (\ref{S^{(2,0)}}) do not receive contributions from $N_{n+1}^{(2)}$, we need to analyze the gauge invariance of them explicitly in the momentum space, which is extremely complicated and we will not attempt to do it in the present work.

\subsection{Consistency checks}
\label{consistency_checks}
\begin{itemize}
    \item The leading correction to the flat space soft graviton theorem, Eq. (\ref{S^{(1,0)}_final}), is invariant under the gauge transformation $\varepsilon_{\alpha\beta}\to\varepsilon_{\alpha\beta}-i\xi_\alpha k_\beta - i\xi_\beta k_\alpha$. Since the gauge transformation at $\mathcal{O}(l^{-2})$ becomes very complicated in the momentum space, to check the gauge invariance of the subleading and sub-subleading corrections to the flat space soft factor, we need to analyze whether Eq. (\ref{LSZ_final}), from which they are derived, is gauge-invariant.
    Since Eq. (\ref{LSZ_final}) is gauge invariant, the subleading and sub-subleading corrections to the flat space soft factor are also gauge invariant.
    \item To check the universal feature of the leading correction term, we have used a different set of scalar field modes (\ref{hp_modes}). The new set of modes has no terms of $\mathcal{O}(l^{-1})$, so there are no leading corrections to the flat space soft factor for this set of modes. Since the leading correction to the soft factor depends on the choice of modes, it is non-universal.
    \item We have used a different set of scalar field modes (\ref{hp_modes}) to compute subleading and sub-subleading corrections to the flat space soft factor. The detailed calculation is given in the Appendix (\ref{Appendix-A}). We have observed that the subleading and sub-subleading corrections for this set of modes are exactly the same as Eqs. (\ref{S^{(2,-1)}}) and (\ref{S^{(2,0)}}). Further, these correction terms only depend on the momenta of the hard particles, soft energy, and the coupling constant.
    Thus, we expect that the subleading and sub-subleading corrections to the flat space soft factor should be universal.
\end{itemize}

\section{Discussions}
\label{discussion_sec}

\begin{table}
\noindent{\scriptsize
\begin{tabular}{cc}
\\
\hline
\vspace{-2pt}&\vspace{-2pt}\\
Flat space soft factor & $\sum\limits_{i=1}^{n} \dfrac{\kappa}{2} \dfrac{p_i^\alpha p_i^\beta\varepsilon_{\alpha\beta}}{\omega p_i\cdot \hat{k}} -i \sum\limits_{i=1}^{n} \dfrac{\kappa}{2}\dfrac{p_i^\alpha\varepsilon_{\alpha\beta}\hat{k}_\gamma J^{\beta\gamma}}{p_i\cdot \hat{k}}$ \\
\vspace{-2pt}&\vspace{-2pt}\\
Leading corrections & 
$-m \sum\limits_{i=1}^n \dfrac{\kappa}{2} \dfrac{p_i^\alpha\varepsilon_{\alpha\beta}\hat{k}_\gamma J^{\beta\gamma}}{(p_i\cdot \hat{k})^2}$ \\
\vspace{-2pt}&\vspace{-2pt}\\
Subleading corrections & $\dfrac{3m^2}{2} \sum\limits_{i=1}^{n} \dfrac{\kappa}{2} \dfrac{p_i^\alpha p_i^\beta \varepsilon_{\alpha\beta}}{\omega(p_i\cdot \hat{k})^3}$ \\
\vspace{-2pt}&\vspace{-2pt}\\
Sub-subleading corrections & $\sum\limits_{i=1}^{n} \dfrac{\kappa}{2} \bigg(
\dfrac{p_i^\alpha p_i^\beta\varepsilon_{\alpha\beta}}{2(p_i\cdot \hat{k})^2} + \dfrac{m^2(p_i^\alpha p_i^\beta \varepsilon_{\alpha\beta})}{(p_i\cdot \hat{k})^3} \hat{k}\cdot\partial_{p_i} + \dfrac{3(p_i^\alpha p_i^\beta\varepsilon_{\alpha\beta})}{2(p_i\cdot \hat{k})^2}p_i\cdot \partial_{p_i} + \dfrac{m^2(p_i^\alpha \varepsilon_{\alpha\beta})}{2(p_i\cdot \hat{k})^2}\partial_{p_i}^\beta \bigg)$ \\
\vspace{-2pt}&\vspace{-2pt}\\
\hline
\end{tabular}}
\caption{Leading, subleading, and sub-subleading corrections to the flat space soft factor.}
\label{table1}
\end{table}

In this paper, we obtained leading, subleading and sub-subleading order corrections induced by the de Sitter geometry in the large curvature length (or small cosmological constant) limit to the flat space soft graviton theorem, and discussed about the universal feature of these corrections. We first obtained mode solutions of the linearized field equation in the background de Sitter space. We then defined the inner product for these modes and verified that they are orthogonal. By defining the perturbative $\mathcal{S}$-matrix in a compact region inside the static patch, and using the LSZ formula, we derived the leading, subleading, and sub-subleading corrections to the flat space soft factor. Further, consistency checks have been performed on these corrections. Performing the gauge transformations, we have verified that the leading, subleading, and sub-subleading corrections to the flat space soft factor are gauge-invariant. To comment on the universality of these corrections, we used a different set of scalar field modes to recompute the leading, subleading, and sub-subleading corrections. We observed that the leading order corrections depend on the choice of modes while the subleading and sub-subleading order corrections remain the same. Thus, the leading corrections to the soft factor are non-universal, while we expect that the subleading and sub-subleading corrections are universal, which is consistent with the results obtained for soft photon theorems in \cite{Sayali_Diksha}. The leading, subleading, and sub-subleading corrections to the flat space soft graviton theorems for the tree-level scattering process are summarized in table (\ref{table1}).

If the subleading and sub-subleading corrections turn out to be universal, it would be interesting to investigate whether perturbative corrections to the flat space asymptotic charges can reproduce these corrections from Ward identities. It is natural to expect such perturbative corrections should start at $\mathcal{O}(l^{-2})$. Recently, the soft graviton theorem in the de Sitter space was derived from Ward identities using the near-horizon supertranslation charges \cite{Mao_Zhou}. Their analysis was done near to the horizon and it is not close to the small compact region where our setup is confined. For the validity of our results, one can not extend the finite boundaries of this region to the cosmological horizons. Also, taking the small cosmological constant limit of the results of \cite{Mao_Zhou} pushes the cosmological horizons to the null infinities in the Minkowski space. Thus, their results can not be expanded naively in $\mathcal{O}(1/l)$ and directly reduce to the flat space soft graviton theorems. In addition to this, the authors in \cite{Mao_Zhou} considered the massless scalars, and the massless limit of scalars is not well defined in our analysis especially for the scalar field modes. Thus, it is not immediately clear to us how our results and those of \cite{Mao_Zhou} can be related.

Moreover, one can study the radiative component of the low frequency gravitational waveforms in the de Sitter background in the small cosmological constant limit to derive the perturbative corrections to the classical soft graviton theorems. One can also evaluate the perturbative corrections to the trajectories of hard particles, and then take the classical limit of our results by replacing all the operators with their classical counterparts. Evaluating perturbative corrections to the classical soft graviton theorem and interpreting various correction terms is an interesting future direction.

In the pursuit of understanding gravitational radiation and memory effects in the presence of a tiny cosmological constant, the analysis presented here might be useful. However, to establish a relation between the soft factors deduced here as a perturbative expansion in cosmological constant, and the (asymptotic) symmetries of the static patch of the de Sitter spacetime, one needs to first analyze the symmetries of the patch of the full de Sitter spacetime and the corresponding charges. Since the static patch only preserves a 4 dimensional subset of the full 10 dimensional de Sitter symmetries, the scattering amplitudes should also obey the subset of symmetries only \cite{Ashtekar_Bonga, Emil_2021}. It would be interesting to see if a relation between the asymptotic charges within this patch, and the soft factors can be established by performing a suitable expansion of the charges around the flat space. This would also be necessary to understand the memory effects in this framework \cite{Compere_Hoque}. Finally, one needs to explore the connection between the asymptotic symmetry group of full de Sitter spacetime, the $\Lambda$-$\text{BMS}_4$ group \cite{Compere_Ruzziconi_2019, Compere_Ruzziconi_2020} and the symmetries of the soft amplitudes considered here.

\section*{Acknowledgements}
D.N.S. and P.C. would like to thank Sayali Bhatkar and Diksha Jain for useful discussions during the initial stages of this work. D.N.S. would also like to thank Alok Laddha for useful discussions and his valuable suggestions. S. B. acknowledges support by DST-SERB/ANRF through MATRICS grant MTR/2022/000170.

\appendix
\section{Perturbative corrections for alternate set of scalar modes}
\label{Appendix-A}
In this section, we use the alternate set of scalar field modes (\ref{hp_modes}) to compute the perturbative corrections to the flat space soft factor. Even though these modes and the corresponding states are non-orthogonal, it is very interesting to study the soft limit of the $\mathcal{S}$-matrix for this set of modes. The $\mathcal{O}(l^{-2})$ corrections to the $\mathcal{S}$-matrix for this new set of modes is given as follows:

\begin{align}
&\Gamma_{n+1}^{(2)}(\{p_1,...,p_n\},k) = -i\sum_{i=1}^{n}\kappa \int d^4z d^4z' \int\dfrac{d^4p}{(2\pi)^4} \bigg(
\dfrac{4m^4(p^\alpha p_i^\beta \varepsilon_{\alpha\beta})}{(p^2+m^2)^4} + \dfrac{2m^2(p^\alpha p_i^\beta \varepsilon_{\alpha\beta})}{(p^2+m^2)^3} \nonumber \\
&+  \dfrac{2(p^\alpha p_i^\beta \varepsilon_{\alpha\beta})}{(p^2+m^2)^2} + \dfrac{3ic (p^\alpha p_i^\beta \varepsilon_{\alpha\beta})}{2m(p^2+m^2)} + \dfrac{5c(p_i\cdot z')(p^\alpha p_i^\beta \varepsilon_{\alpha\beta})}{2m(p^2+m^2)} - \dfrac{(p_i\cdot z')^2(p^\alpha p_i^\beta \varepsilon_{\alpha\beta})}{2m^2(p^2+m^2)} \nonumber \\
&-\dfrac{ic(p_i\cdot z')^2(p^\alpha p_i^\beta \varepsilon_{\alpha\beta})}{2m(p^2+m^2)}
+ \dfrac{i(p_i\cdot z')^3(p^\alpha p_i^\beta \varepsilon_{\alpha\beta})}{6m^2(p^2+m^2)} +\dfrac{z'^2(p^\alpha p_i^\beta \varepsilon_{\alpha\beta})}{4(p^2+m^2)} - \dfrac{icmz'^2(p^\alpha p_i^\beta \varepsilon_{\alpha\beta})}{2(p^2+m^2)} \nonumber \\
&+ \dfrac{iz'^2(p_i\cdot z')(p^\alpha p_i^\beta \varepsilon_{\alpha\beta})}{4(p^2+m^2)} - \dfrac{2im^2(p\cdot z)(p^\alpha p_i^\beta \varepsilon_{\alpha\beta})}{(p^2+m^2)^3}
- \dfrac{i(p\cdot z)(p^\alpha p_i^\beta \varepsilon_{\alpha\beta})}{(p^2+m^2)^2}
+ \dfrac{m^2 z^2 (p^\alpha p_i^\beta \varepsilon_{\alpha\beta})}{2(p^2+m^2)^2} \nonumber \\
&- \dfrac{z^2(p^\alpha p_i^\beta \varepsilon_{\alpha\beta})}{2(p^2+m^2)}
+ \dfrac{3i(p^\alpha z'^\beta \varepsilon_{\alpha\beta})}{2(p^2+m^2)}
+ \dfrac{cm(p^\alpha z'^\beta \varepsilon_{\alpha\beta})}{p^2+m^2}
+ \dfrac{(p\cdot z')(p_i^\alpha z'^\beta \varepsilon_{\alpha\beta})}{2(p^2+m^2)} 
- \dfrac{(p\cdot p_i) (z'^\alpha z'^\beta \varepsilon_{\alpha\beta})}{4(p^2+m^2)} \bigg) \nonumber \\
&\cdot e^{i(p-p_i-k)\cdot z'} e^{-ip\cdot z} \int\prod_{j=1; j\ne i}^{n-1} d^4z_j \sqrt{-g(z_j)} g^*_{p_j}(z_j)(-i)(\nabla^{2}_{z_j}-m^2) \big\langle G[\{\phi(z_j) \}] \big\rangle.
\end{align}
To simplify the above expression, we first replace all $z'$ by $-i\partial_p$, and then perform the integration by parts in $p$. Next, we perform the $z'$-integral which yields the delta function. We finally integrate this delta function and take the soft limit and obtain
\begin{align}
\label{Append_Gamma_soft}
&\Gamma_{n+1}^{(2)}(\{p_1,...,p_n\},k) = -i\sum_{i=1}^{n} \frac{\kappa}{2} \int d^4z \bigg(
\dfrac{3m^2(p_i^\alpha p_i^\beta\varepsilon_{\alpha\beta})}{2(p_i\cdot k)^3}
+\dfrac{p_i^\alpha p_i^\beta\varepsilon_{\alpha\beta}}{2(p_i\cdot k)^2}
+\dfrac{3ic(p_i^\alpha p_i^\beta\varepsilon_{\alpha\beta})}{2m(p_i\cdot k)} \nonumber \\
&+ \dfrac{im^2(p_i^\alpha p_i^\beta\varepsilon_{\alpha\beta})(k\cdot z)}{2(p_i\cdot k)^3}
- \dfrac{2i(p_i^\alpha p_i^\beta\varepsilon_{\alpha\beta})(k\cdot z)}{(p_i\cdot k)^2}
- \dfrac{cm(p_i^\alpha p_i^\beta\varepsilon_{\alpha\beta})(k\cdot z)}{(p_i\cdot k)^2}
- \dfrac{3i(p_i^\alpha p_i^\beta \varepsilon_{\alpha\beta})(p_i\cdot z)}{2(p_i\cdot k)^2} \nonumber \\
&+ \dfrac{5c(p_i^\alpha p_i^\beta \varepsilon_{\alpha\beta})(p_i\cdot z)}{2m(p_i\cdot k)}
+ \dfrac{(p_i^\alpha p_i^\beta \varepsilon_{\alpha\beta})(p_i\cdot z)(k\cdot z)}{2(p_i\cdot k)^2}
- \dfrac{(p_i^\alpha p_i^\beta\varepsilon_{\alpha\beta})(p_i\cdot z)^2}{2m^2(p_i\cdot k)} 
- \dfrac{ic(p_i^\alpha p_i^\beta\varepsilon_{\alpha\beta})(p_i\cdot z)^2}{2m(p_i\cdot k)} \nonumber \\
&+ \dfrac{i(p_i^\alpha p_i^\beta\varepsilon_{\alpha\beta})(p_i\cdot z)^3}{6m^2(p_i\cdot k)} 
- \dfrac{z^2 (p_i^\alpha p_i^\beta \varepsilon_{\alpha\beta})}{4(p_i\cdot k)}
- \dfrac{icm z^2 (p_i^\alpha p_i^\beta \varepsilon_{\alpha\beta})}{2(p_i\cdot k)}
+ \dfrac{i z^2 (p_i^\alpha p_i^\beta \varepsilon_{\alpha\beta}) (p_i\cdot z)}{4(p_i\cdot k)}
- \dfrac{im^2(p_i^\alpha z^\beta\varepsilon_{\alpha\beta})}{2(p_i\cdot k)^2} \nonumber \\
&+ \dfrac{4i(p_i^\alpha z^\beta\varepsilon_{\alpha\beta})}{p_i\cdot k}
+ \dfrac{2cm(p_i^\alpha z^\beta\varepsilon_{\alpha\beta})}{p_i\cdot k}
+ \dfrac{(p_i^\alpha z^\beta\varepsilon_{\alpha\beta})(k\cdot z)}{2(p_i\cdot k)}
+ \dfrac{m^2(z^\alpha z^\beta \varepsilon_{\alpha\beta})}{4(p_i\cdot k)} 
- \dfrac{z^\alpha z^\beta \varepsilon_{\alpha\beta}}{4}
\bigg) \nonumber \\ 
&\bigg(1-ik\cdot z -\dfrac{1}{2} (k\cdot z)^2 + ... \bigg) e^{-ip_i \cdot z}
\int\prod_{j=1; j\ne i}^{n-1} d^4z_j \sqrt{-g(z_j)} g^*_{p_j}(z_j)(-i)(\nabla^{2}_{z_j}-m^2) \big\langle G[\{\phi(z_j) \}] \big\rangle.
\end{align}
By identifying the terms correspond to $A^{(0)}$ and $A^{(1)}$ in the above expression, and separating them from the rest of the terms, we can rewrite Eq. (\ref{Append_Gamma_soft}) as
\begin{align}
&\Gamma_{n+1}^{(2)}(\{p_1,...,p_n\},k) = -i\sum_{i=1}^{n} \frac{\kappa}{2} \int d^4z \bigg(
\dfrac{3m^2(p_i^\alpha p_i^\beta\varepsilon_{\alpha\beta})}{2(p_i\cdot k)^3}
+\dfrac{p_i^\alpha p_i^\beta\varepsilon_{\alpha\beta}}{2(p_i\cdot k)^2}
- \dfrac{im^2(p_i^\alpha p_i^\beta\varepsilon_{\alpha\beta})(k\cdot z)}{(p_i\cdot k)^3} \nonumber \\
&- \dfrac{3i(p_i^\alpha p_i^\beta \varepsilon_{\alpha\beta})(p_i\cdot z)}{2(p_i\cdot k)^2}
- \dfrac{im^2(p_i^\alpha z^\beta\varepsilon_{\alpha\beta})}{2(p_i\cdot k)^2} + \mathcal{O}(k^{-1}) \bigg) e^{-ip_i \cdot z} \nonumber \\
&\cdot\int\prod_{j=1; j\ne i}^{n-1} d^4z_j e^{-ip_j\cdot z_j}(-i)(\nabla^{2}_{z_j}-m^2) \big\langle G[\{\phi(z_j) \}] \big\rangle
+ A^{(1)} \Gamma_n^{(1)} + A^{(0)} \Gamma_n^{(2)},
\end{align}
where $\mathcal{O}(k^{-1})$ terms contribute beyond the sub-subleading order and are suppressed.
Now, we replace all $z$ by $i\partial_{p_i}$ in the above expression and put the $\mathcal{O}(l^{-2})$ corrections to the flat space soft factor in the momentum space as follows:
\begin{align}
A^{(2)} = \sum_{i=1}^{n} \frac{\kappa}{2} \bigg(
&\dfrac{3m^2(p_i^\alpha p_i^\beta\varepsilon_{\alpha\beta})}{2(p_i\cdot k)^3} + \dfrac{p_i^\alpha p_i^\beta\varepsilon_{\alpha\beta}}{2(p_i\cdot k)^2} + \dfrac{m^2(p_i^\alpha p_i^\beta \varepsilon_{\alpha\beta})}{(p_i\cdot k)^3} k\cdot\partial_{p_i} + \dfrac{3(p_i^\alpha p_i^\beta\varepsilon_{\alpha\beta})}{2(p_i\cdot k)^2}p_i\cdot \partial_{p_i} \nonumber \\ 
&+ \dfrac{m^2(p_i^\alpha \varepsilon_{\alpha\beta})}{2(p_i\cdot k)^2}\partial_{p_i}^\beta \bigg) + \mathcal{O}(k^{-1}).
\end{align}
We must note that it is independent of the arbitrary constant $c$ present in the scalar field modes in (\ref{hp_modes}). The subleading and sub-subleading corrections to the flat space soft factor for this new set of modes (\ref{hp_modes}) are
\begin{equation}
S^{(2,-1)} = \sum_{i=1}^{n} \frac{\kappa}{2} \dfrac{3m^2(p_i^\alpha p_i^\beta \varepsilon_{\alpha\beta})}{2\omega(p_i\cdot \hat{k})^3},
\end{equation}
\begin{equation}
S^{(2,0)} = \sum_{i=1}^{n} \frac{\kappa}{2} \bigg(
\dfrac{p_i^\alpha p_i^\beta\varepsilon_{\alpha\beta}}{2(p_i\cdot \hat{k})^2} + \dfrac{m^2(p_i^\alpha p_i^\beta \varepsilon_{\alpha\beta})}{(p_i\cdot \hat{k})^3} \hat{k}\cdot\partial_{p_i} + \dfrac{3(p_i^\alpha p_i^\beta\varepsilon_{\alpha\beta})}{2(p_i\cdot \hat{k})^2}p_i\cdot \partial_{p_i} + \dfrac{m^2(p_i^\alpha \varepsilon_{\alpha\beta})}{2(p_i\cdot \hat{k})^2}\partial_{p_i}^\beta \bigg).
\end{equation}

\end{document}